\documentclass[a4paper,11pt]{article}
\usepackage[utf8]{inputenc}
\usepackage{authblk}
\usepackage{natbib}
\usepackage{amssymb,amsmath}
\usepackage[a4paper, total={6.5in, 9in}]{geometry}
\usepackage{graphicx}
\usepackage{listings}
\usepackage{epstopdf}
\usepackage{courier}
\usepackage{tabularx}
\usepackage{hyperref}

\begin{document}

\title{\vspace{-1.0cm}Probabilistic modeling of occurring substitutions in PAR-CLIP data }

\author[1,2,3]{Monica Golumbeanu}
\author[1,2,3]{Pejman Mohammadi}
\author[1,2,4]{Niko Beerenwinkel}

\affil[1]{\normalsize Department of Biosystems Science and Engineering, ETH Zürich, Basel, Switzerland}
\affil[2]{\normalsize SIB Swiss Institute of Bioinformatics, Basel, Switzerland}
\affil[3]{\normalsize Equal contribution}
\affil[4]{\normalsize E-mail \url{niko.beerenwinkel@bsse.ethz.ch}}


\date{}
\maketitle

\begin{abstract}
\normalsize
\begin{sloppypar}
Photoactivatable ribonucleoside-enhanced cross-linking and immunoprecipitation (PAR-CLIP) is an experimental method based on next-generation sequencing for identifying the RNA interaction sites of a given protein. The method deliberately inserts T-to-C substitutions at the RNA-protein interaction sites, which provides a second layer of evidence compared to other CLIP methods. However, the experiment includes several sources of noise which cause both low-frequency errors and spurious high-frequency alterations. Therefore, rigorous statistical analysis is required in order to separate true T-to-C base changes, following cross-linking, from noise. So far, most of the existing PAR-CLIP data analysis methods focus on discarding the low-frequency errors and rely on high-frequency substitutions to report binding sites, not taking into account the possibility of high-frequency false positive substitutions. Here, we introduce~\emph{BMix}, a new probabilistic method which explicitly accounts for the sources of noise in PAR-CLIP data and distinguishes cross-link induced T-to-C substitutions from low and high-frequency erroneous alterations. We demonstrate the superior speed and accuracy of our method compared to existing approaches on both simulated and real, publicly available human datasets. The model is implemented in the Matlab toolbox~\emph{BMix}, freely available at \href{www.cbg.bsse.ethz.ch/software/BMix}{www.cbg.bsse.ethz.ch/software/BMix}.
\end{sloppypar}
\end{abstract}

\section{Introduction}

\begin{figure*}[!tpb]
	\centerline{\includegraphics[scale=0.37]{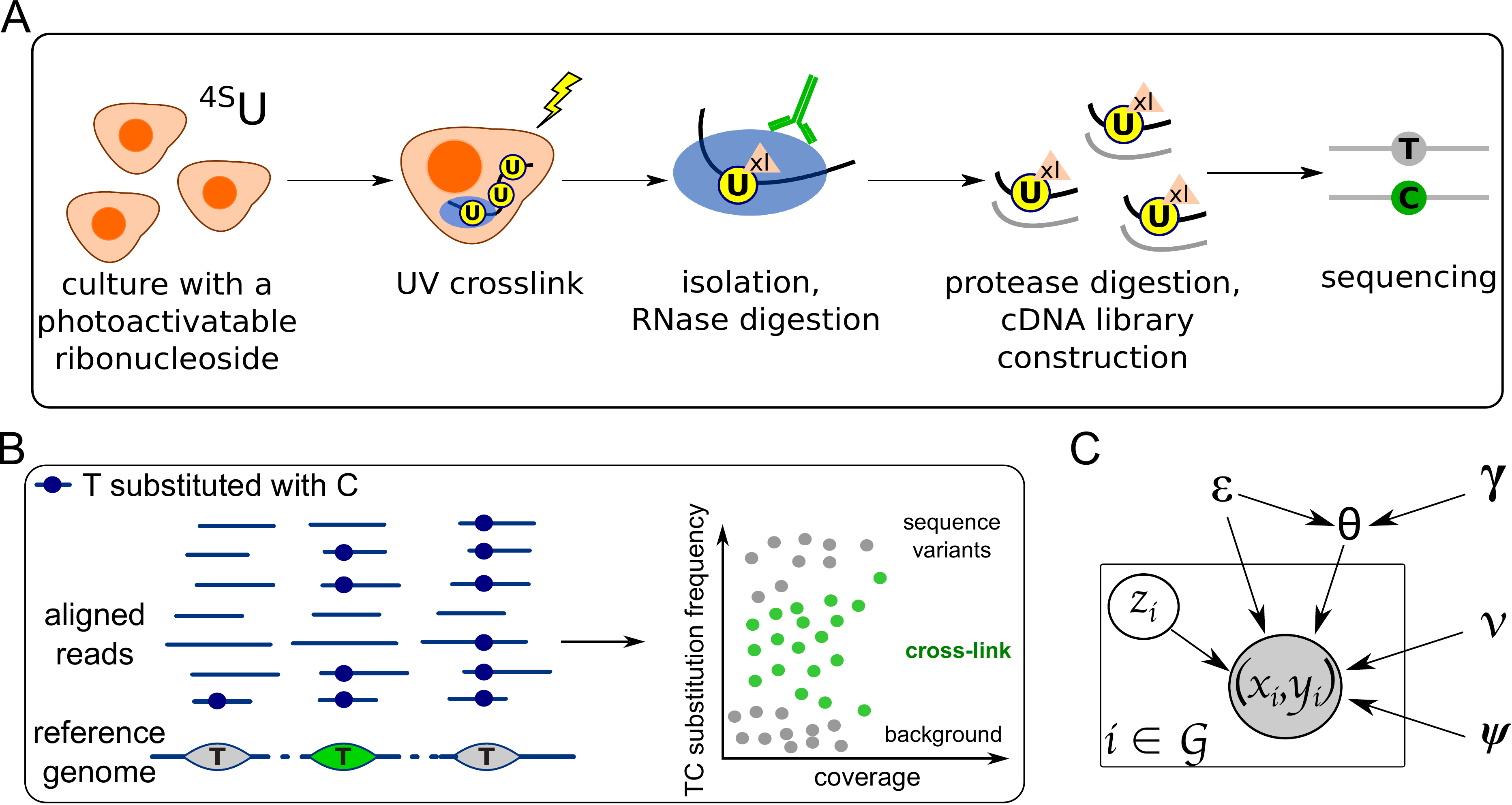}}
	\caption{~\textbf{(A)} The main steps of the PAR-CLIP protocol including cell culture with 4-thiouridine ($^{4S}U$), UV cross-link, isolation of the RNA-RBP complexes and sequencing of the bound RNA fragments where systematic T-to-C substitutions occur. \textbf{(B)}~Schematic representation of the BMix rationale. The method takes as input the aligned sequencing reads and uses a three-component mixture based on the observed substitution counts to infer whether the observed T to C substitutions are most likely caused by sequencing errors, sequence variants or cross-link. \textbf{(C)}~Graphical representation of the statistical model. For each evaluated locus $i$ on the genome $\mathcal{G}$, the coverage $x_i$ and the mismatch count $y_i$ are observed. The latent variable $z_i$ used to indicate the different components of the mixture, and the parameters $\epsilon, \gamma, \theta, \nu$, and $\psi$ define the model (cf. Methods).
	}
	\label{figure1}
\end{figure*}

\begin{sloppypar}
\noindent RNA molecules interact with proteins and form ribonucleoprotein complexes (RNPs) actively involved in a plethora of essential biological processes such as translational regulation, alternative splicing or RNA transport~\citep{lunde, muller}. A well-known example of RNA-binding proteins (RBPs) consists of members of the Argonaute family, components of the RNA-induced silencing complex (RISC), which bind to diverse small RNA molecules and regulate gene silencing~\citep{Meister2013}. Gerstberger~\emph{et. al.}~report 1,542 RBPs in humans~\citep{Gerstberger2014}, many of these having been found dysregulated in diseases including cancer~\citep{Lukong2008, Kechavarzi2014}. Therefore, characterizing the interactions between RNA and RBPs represents an important step towards understanding RNA function.
\end{sloppypar}
\begin{sloppypar}
\indent High-throughput sequencing technology allows querying the binding sites of a specific RBP in a transcriptome-wide fashion~\citep{Blencowe2009, Kloetgen2014}. One of the recently developed high-throughput sequencing-based experimental protocols aiming to identify the binding sites of RBPs throughout the transcriptome is Photo-Activatable Ribonucleoside-enhanced
Crosslinking and Immunoprecipitation (PAR-CLIP)~\citep{Hafner2010}. According to this method~(Figure~\ref{figure1}A), a synthetic photoactivatable ribonucleoside such as $^{4S}U$ (4-thiouridine) or, less commonly, $^{6S}G$ (6-thioguanosine) is integrated into the RNA of cultured cells. Upon exposure to ultraviolet (UV) light, cross-linking of~RBPs to RNA occurs. The cross-linked RNA-RBP pairs are subsequently isolated using immunoprecipitation with an antibody targeting the protein of interest, and the RNA fragment is retrieved upon protein digestion. A complementary DNA (cDNA) sequencing library is generated by reverse complementing the RNA fragments. Due to the incorporated nucleoside, systematic T-to-C (for $^{4S}U$) or G-to-A (for  $^{6S}G$) substitutions appear in the cDNA library at the interaction sites~\citep{Hafner2010}. Therefore, PAR-CLIP brings the advantage of having an additional layer of evidence by introducing specific base changes at the binding sites. 
\end{sloppypar}

\indent PAR-CLIP data is characterized by prevalent T-to-C substitutions observed at different mismatch frequencies~\citep{Hafner2010}. However, compared to RNA-Seq, PAR-CLIP has been observed to also introduce a large number of other substitutions, different from T-to-C, notably at low and high mismatch frequency (see Results). This indicates the presence of noise and contamination in the PAR-CLIP procedure which can as well introduce erroneous T-to-C substitutions, with high potential to be mistaken for true cross-link alterations. Therefore, of great importance in PAR-CLIP data analysis is discarding the low-frequency sequencing errors, as well as high-frequency substitutions induced by contamination or sequence variants.\\
\indent Currently, there is a handful of methods available for analyzing PAR-CLIP data, trying to identify the RNA-RBP binding sites using various techniques, such as kernel density estimation~\citep{corcoran}, non-parametric mixture models~\citep{cem, comoglio2015}, Bayesian hidden Markov models~\citep{Yun2014}, and binomial tests~\citep{Chen2014}. Some of the available methods focus on analyzing data from one specific type of RBP, such as, for example, AGO2 PAR-CLIP data~\citep{Erhard2013}. However, the non-cross-link, high-frequency substitutions are usually reported within high-confidence binding sites by most of these methods. To the best of our knowledge, only one of the currently existing PAR-CLIP analysis methods, namely WavClusteR~\citep{cem,comoglio2015}, accounts for these substitutions. Sievers~\emph{et.al.} show that cross-link loci reside in moderately altered sites~\citep{cem}, and, within the WavClusteR package, use a mixture model based on the relative substitution frequencies to exclude sites with too low or high nucleotide substitution rates. However, the method is not widely used due to complex implementation and long execution time. Additionally, the approach underlying the WavClusteR package does not explicitly model read counts within the genome and eventually uses a fixed cutoff of the substitution frequencies to select high-confidence T-to-C alterations, which leads to a higher error rate, especially in low-coverage regions.\\
\indent Here, we present a novel probabilistic approach, based on a constrained three-component binomial mixture, to explicitly describe substitution counts observed in PAR-CLIP data. The method, coined BMix, uses a semi-supervised maximum likelihood approach to estimate the substitution rates induced by PAR-CLIP and separates both low and high-frequency erroneous T-to-C alterations from the true cross-link substitutions (Figure~\ref{figure1}B). A toolbox containing the entire PAR-CLIP analysis pipeline employing the presented method is available for download at~\href{www.cbg.bsse.ethz.ch/software/BMix}{www.cbg.bsse.ethz.ch/software/BMix}. We perform an exhaustive comparison of our method to existing approaches and demonstrate the increased performance of BMix in terms of speed, accuracy, and consistency on synthetic and real PAR-CLIP data.

~
\section{Methods}
\subsection{Data preprocessing}
\begin{sloppypar}
	PAR-CLIP and RNA-Seq reads were clipped from their adapter using the~\texttt{fastx clipper} tool~(\href{http://hannonlab.cshl.edu}{http://hannonlab.cshl.edu}) and only reads larger than 13 nucleotides were kept for further analysis. The reads were aligned to the human reference genome assembly~\emph{hg19} with the~\texttt{bowtie} alignment tool version 0.12.9~\citep{bowtie} using the same parameters employed by PARalyzer and WavClusteR,~\texttt{-n 1 --best -m 100 -k 1 -l 50}.
\end{sloppypar}
\subsection{Probabilistic modeling of substitution counts}
The data consists of PAR-CLIP cDNA sequencing reads. After alignment to the reference genome, at each position in an aligned read, the observed nucleotide can either match or differ from the reference. When the nucleotide differs from the reference, several causes are possible. First, the observed nucleotide could be a sequence variant, caused by single-nucleotide polymorphism (SNP), or foreign, non-cross-linked RNA fragments mapped to a reference location highly similar to their sequence (contamination). Second, if the reference is T and the corresponding read nucleotide is C, then a cross-link-induced substitution could have occurred. Third, a mismatch could have also happened due to sequencing error. These three events are not mutually exclusive.\\
\indent In order to detect RNA-protein cross-link-induced T-to-C substitutions, we model, for each position $i$ in the genome, where the reference, $r_i$, is different from C, the probability of the observed T-to-C, A-to-C, or G-to-C substitution. We define $x_i$ as the sequencing coverage at position $i$, and $y_i$ as the number of times the reference nucleotide is substituted with C in all the reads covering position $i$. We assume that the observed T-to-C alterations are due to (i) sequencing error, (ii) SNPs or contamination, or (iii) PAR-CLIP cross-link-based substitution, whereas the observed A-to-C and G-to-C substitutions are assumed to originate only from (i) sequencing error, or (ii) SNPs or contamination. We ignore the cases of photo-activated sequence variants (i.e., where the reference is A, or G, while the sequence variant is T, and is substituted to C due to photo-activation by PAR-CLIP). With this simplification, we introduce the latent random variable $z_i \in \{1,2,3\}$ corresponding to the three possible reasons (i)-(iii) that can explain the observed nucleotide at locus $i$ on the genome. Specifically, for reference T positions, $z_i=1$ refers to background, $z_i=2$ corresponds to a sequence variant, and $z_i=3$ refers to an RNA-RBP cross-link. For reference A or G positions, only $z_i=1$ and $z_i=2$ are possible.\\
\indent We define $\epsilon$ as the probability of inducing a substitution due to sequencing noise. This probability accounts for all the modeled nucleotide substitutions (i.e., T-to-C, A-to-C, G-to-C, C-to-T, C-to-A, and C-to-G) and is expected to be low. Consequently, at background positions ($z_i=1$), the probability of occurrence of a specific substitution is $\epsilon$. Furthermore, $3\epsilon$ represents the probability of a nucleotide to mutate to any possible base. Therefore, in the case of sequence variant loci ($z_i=2$), where one can assume that the aligned reads originate from a different sequence than the reference, substitutions happen with a success probability of $1-3\epsilon$. Finally, at cross-link loci ($z_i=3$), which can, at the same time, be affected by sequencing errors, T-to-C substitutions occur with probability
\begin{equation}
\theta = (1-\gamma)\epsilon + (1-3\epsilon)\gamma,
\end{equation} 
where $\gamma$ corresponds to the probability of a T nucleotide to be mutated to C following photo-activation and cross-link during PAR-CLIP, i.e., to the efficiency of the protocol to induce T-to-C substitutions at cross-link loci. We assume that the probability $\theta$ is bounded between $\epsilon$ and $1-3\epsilon$, which results in the constraint $\epsilon \leq 0.25$.\\
\indent We denote by $\nu = P(z_i=2)$ the probability of a locus on the genome to be a sequence variant, and, for the remaining cases which are not sequence variants, by $\psi$ the probability that a genomic locus is a cross-link site. The observed data at each T reference position on the genome is then modeled by a constrained mixture of three binomial distributions, and the probability of an observed data point is
\begin{equation}
\begin{split}
P((x_i,y_i) \mid r_i=T) &= \sum_{z_i=1}^3 P((x_i,y_i) \mid z_i, r_i=T)P(z_i \mid r_i=T) \\
&= \underbrace{(1-\psi)(1-\nu) Bin(y_i; x_i, \epsilon)}_{background} + \\
&+ \underbrace{\nu Bin(y_i; x_i, 1-3\epsilon)}_{sequence~variant} + \underbrace{\psi(1-\nu)Bin(y_i; x_i, \theta)}_{cross-link}
\end{split}
\label{tc}
\end{equation} 
where $\epsilon, \theta, \gamma, \nu$, and $\psi$ are the parameters of the model, and the notation $Bin(k;n,p)$ corresponds to the probability mass function of the Binomial distribution, precisely the probability of having $k$ successes within $n$ trials with success probability $p$.\\
\indent In absence of a control PAR-CLIP experiment, our model readily incorporates information from A-to-C and G-to-C alterations for a better estimation of the sequencing error $\epsilon$ and the probability $\nu$. The probability of observed coverage and mismatch count for observed A-to-C and G-to-C alterations is
\begin{equation}
P((x_i, y_i) | r_i \in \{A,G\}) = \underbrace{(1-\nu)Bin(y_i; x_i, \epsilon)}_{background}+\underbrace{\nu Bin(y_i; x_i, 1-3\epsilon)}_{sequence~variant}.
\label{notc}
\end{equation}
\indent The model (Figure~\ref{figure1}C) is thus fully defined by equation~\ref{tc} for reference genome T positions, and equation~\ref{notc} for A and G positions in the reference genome. Using these equations, we can derive the likelihood for the entire set of observations $\mathcal{D} = \{(x_i,y_i)\}_{i\in\mathcal{G}}$ throughout the whole genome $\mathcal{G}$ as follows:
\begin{equation}
L(\epsilon, \theta, \gamma, \nu, \psi) = P(\mathcal{D} \mid \epsilon, \theta, \gamma, \nu, \psi) = \prod_{i \in \mathcal{G}}P((x_i, y_i )\mid r_i)
\end{equation}
\indent We infer all the parameters of our model by maximizing the above defined likelihood with a gradient-based nonlinear constrained optimization~\citep{Powell1978}. To classify each T locus on the genome as either background, sequence variant, or cross-link, we choose the maximum of the posterior probabilities of the latent variable $z_i$:
\begin{equation}
\begin{split}
P(z_i=1 \mid (x_i, y_i), r_i=T) &\propto (1-\psi)(1-\nu) Bin(y_i; x_i, \epsilon)\\
P(z_i=2 \mid (x_i, y_i), r_i=T) &\propto \nu Bin(y_i; x_i, 1-3\epsilon) \\
P(z_i=3 \mid (x_i, y_i), r_i=T) &\propto \psi(1-\nu)Bin(y_i; x_i, \theta) 
\end{split}
\end{equation}
where the $\propto$ symbol is used to represent proportionality between the posterior probability and the product between the likelihood and the prior, thus the normalization constant (same in all three cases) being omitted.


\subsection{Construction of RNA-RBP binding sites}
Once the high-confidence cross-link T loci were identified, we report candidate RNA-RBP binding sites by using the sequencing reads that span these loci. By~\emph{binding site} we denote the region on the transcriptome where the protein of study attaches in order to fulfill a specific function. In order to construct the binding sites, all the reads spanning a cross-link locus are grouped into a cluster. The low-coverage boundaries are trimmed and overlapping clusters (at least 1 nucleotide) are grouped into contigs and reported as candidate RNA-RBP binding sites (Supplementary Figure~S1).
\subsection{Generation of the simulated data}
\indent For the generation of realistic synthetic data, we used the AGO2 PAR-CLIP data published in~\citep{Kishore2011} and we introduced systematic A-to-C substitutions. Precisely, starting from 2000 known random nucleotide A positions on chromosome 1, we introduced an A-to-C base change in each sequencing read with probability $\mu$ at the respective positions. Furthermore, we changed to C, using the same probability, the neighboring A nucleotides within an interval of 50 bases centered in each of the 2000 positions. By altering the neighboring positions, we built regions where the incorporated A-to-C substitutions were more dense, simulating binding sites. The probability $\mu$ denotes how likely an activated nucleotide T within a binding site is mutated to C in the real PAR-CLIP protocol. The produced simulated data contains a realistic amount of sequencing errors and contamination and is based on the alteration of a reference base different from T. We assessed the performance of BMix, PARalyzer and WavClusteR on the simulated data. The T-to-C substitutions were ignored by all the applied methods on the simulated data and did not affect their outcome.
\begin{sloppypar}
	\subsection{Comparison with other methods}
	\indent We compared BMix to PARalyzer v1.1~\citep{corcoran} and WavClusteR v2.0.0~\citep{comoglio2015} on the produced synthetic data, as well as on publicly available PAR-CLIP datasets published in~\citep{Kishore2011} and~\citep{cem}. On the simulated data, the three methods were compared in terms of accuracy. The accuracy is defined as the ratio of true positive and negative substitutions over the entire set of observed substitutions. On real data, the methods were evaluated according to specific characteristics of the studied proteins, such as their affinity for microRNA (miRNA), 3'-untranslated regions (3'UTRs), and introns annotated in the RefSeq database~(\href{http://www.ncbi.nlm.nih.gov/refseq/}{http://www.ncbi.nlm.nih.gov/refseq/}), enrichment of protein-specific RNA recognition elements (RRE), as well as execution time. The reported execution time corresponds to the amount of time spent running by each method on one core of a Linux machine with a clock rate of 2.3GHz.
\end{sloppypar}

\begin{figure}[t]
	\centering
	\includegraphics[width=0.55\textwidth]{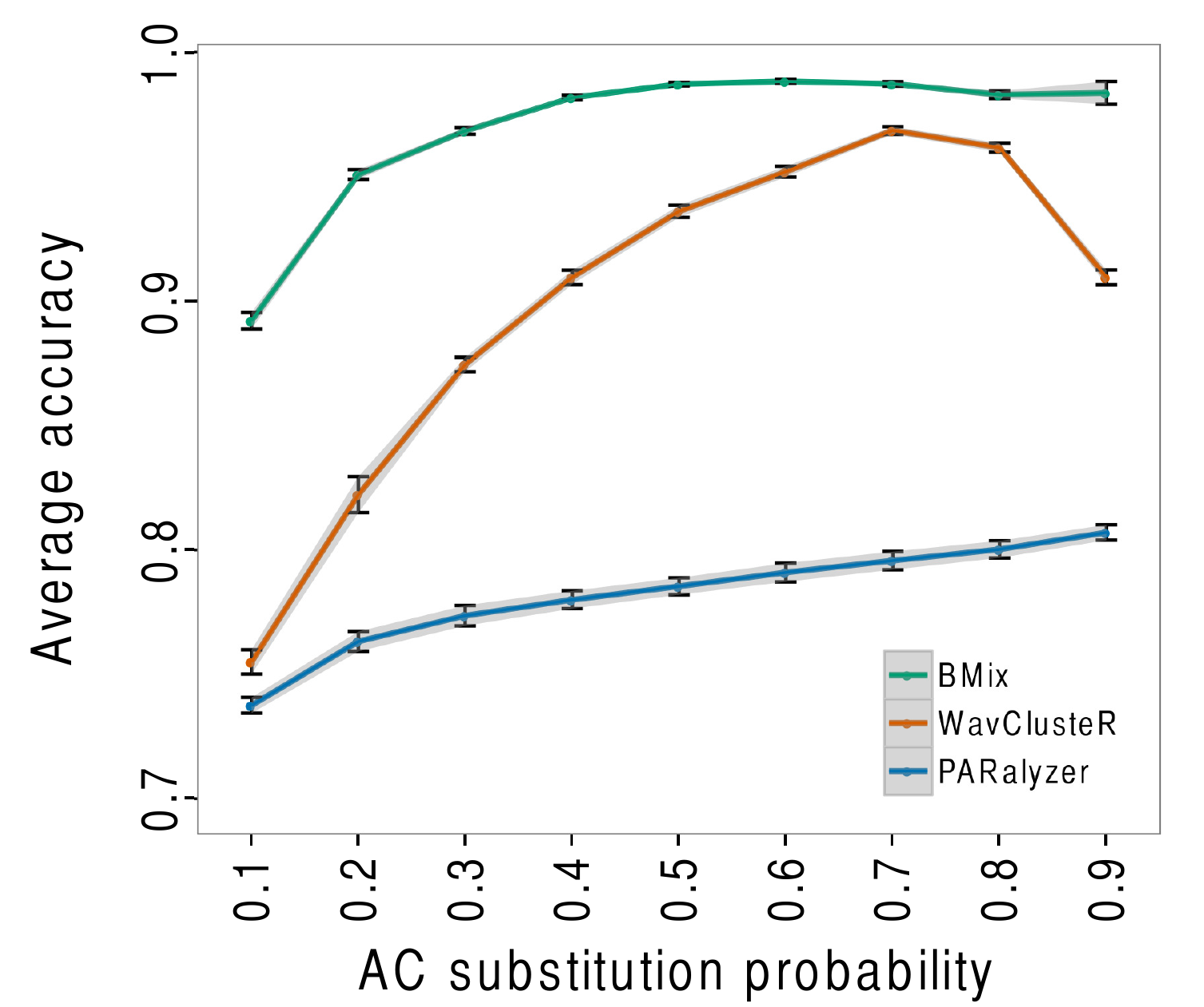}
	\caption{Accuracy of BMix, WavClusteR, and PARalyzer on 100 simulated datasets generated with different substitution probabilities. The error bars and grey shade correspond to one standard deviation of the accuracy over 100 synthetic datasets.}
	\label{accuracy}
\end{figure}

\section{Results}
\noindent We validated our model on synthetic and real human PAR-CLIP data and assessed its performance compared to PARalyzer and WavClusteR methods (see Methods). 
\subsection{Performance on simulated PAR-CLIP data}
In the absence of a ground truth dataset for protein-binding sites, we generated a set of synthetic datasets in order to evaluate our model and compare it to the existing methods WavClusteR and PARalyzer. We started from real world PAR-CLIP data and mimicked~\emph{in silico} the PAR-CLIP protocol for a different substitution than T-to-C, precisely A-to-C. In this way, we kept the intrinsic noise and contamination levels specific to PAR-CLIP data and, at the same time, introduced validation loci. Furthermore, the simulated data was built independently from our model, providing an unbiased test set for all the methods. Knowing thus which A-to-C substitutions were introduced, we could test how accurate BMix as well as other methods were in detecting the altered loci. For each substitution probability $\mu$ in the simulation, ranging from $0.1$ to $0.9$, we generated 100 datasets (cf. Methods) and computed the accuracy of BMix, WavCluster, and PARalyzer.\\
\indent On simulated data, BMix had a significantly higher accuracy (Wilcoxon test $p < 10^{-4}$) than the other two methods, for all the employed substitution probabilities $\mu$ (Figure~\ref{accuracy}). BMix had on average an accuracy of 97\%, compared to 90\% for WavClusteR and 78\% for PARalyzer. All methods reached their lowest accuracy at $\mu=0.1$. However, while WavClusteR and PARalyzer had similar accuracy for this case (76\% and 74\%, respectively), BMix outperformed them with an accuracy of 89\%.\\
\indent By looking at the outcome of the three methods on a randomly chosen synthetic dataset with $\mu=0.5$ (Supplementary Figure S2), the main characteristics of each method were exposed. After learning non-linear classification boundaries, BMix managed to detect 98.5\% of the induced cross-link loci keeping a very low false positive rate of 1\%, and a low false negative rate of 1.5\%. On the other hand, PARalyzer detected only 60\% of the correct loci with a twice as much false positive rate. WavClusteR had the lowest false positive rate at 0.4\%, but a false negative rate of 12.6\%. Furthermore, as expected, PARalyzer especially failed to discard high-frequency altered non-cross-link loci and reported all of them, which explains the increased false positive rate. While WavClusteR successfully managed to eliminate the high-frequency spurious substitutions, the method had difficulty in detecting the cross-link loci in the lower coverage regions leading to an increased false negative rate.

\subsection{Application to real PAR-CLIP datasets}
\begin{figure}[t]
	\centering
	\includegraphics[width=0.8\textwidth]{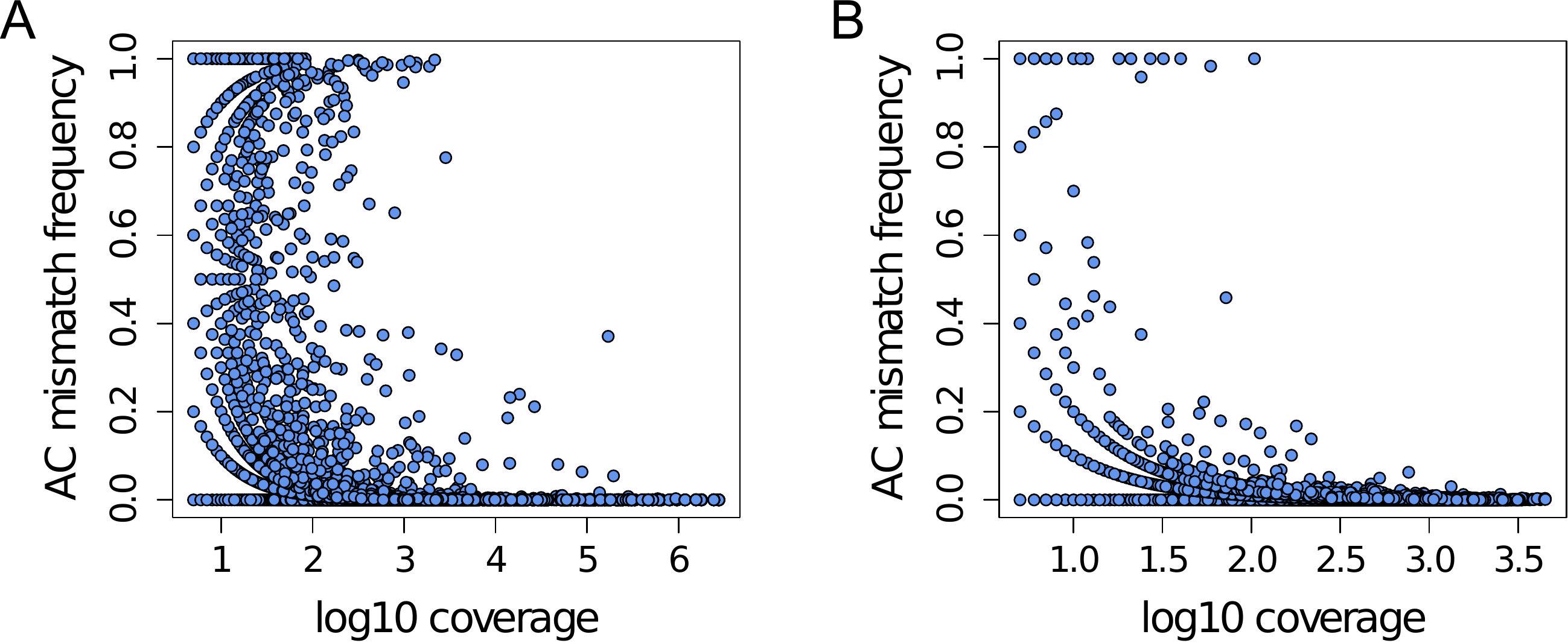}
	\caption{Genome-wide A-to-C observed mismatch frequency as a function of log10 of  coverage in a PAR-CLIP dataset~\textbf{(A)}, as well as in the matched control RNA-Seq dataset~\textbf{(B)}.}
	\label{AC}
\end{figure}

We ultimately applied our method on three published human PAR-CLIP datasets corresponding to proteins AGO2, HUR~\citep{Kishore2011}, and MOV10~\citep{cem}, and inferred the model parameters for each dataset (Supplementary Table 1). Additionally, we applied WavClusteR and PARalyzer on the same data and we compared the three different methods by evaluating their results according to specific characteristics of the proteins such as miRNA, 3'UTR, and intron affinity, as well as enrichment of protein-specific~ RNA recognition elements (RREs).\\
\indent By comparing the PAR-CLIP data for the AGO2 protein to matched RNA-Seq data from the same sample, published in~\citep{Kishore2011}, we observed the expected prevalence of T-to-C substitutions (Supplementary Figure S3), but also a significantly larger amount of A-to-C and G-to-C alterations (one tailed Wilcoxon test $p < 10^{-3}$) in the AGO2 PAR-CLIP dataset (Figure~\ref{AC} and Supplementary Figure S4) indicating a high level of contamination.
\vspace{0.2cm}

\begin{figure*}[th!]
	\centerline{\includegraphics[width=\textwidth]{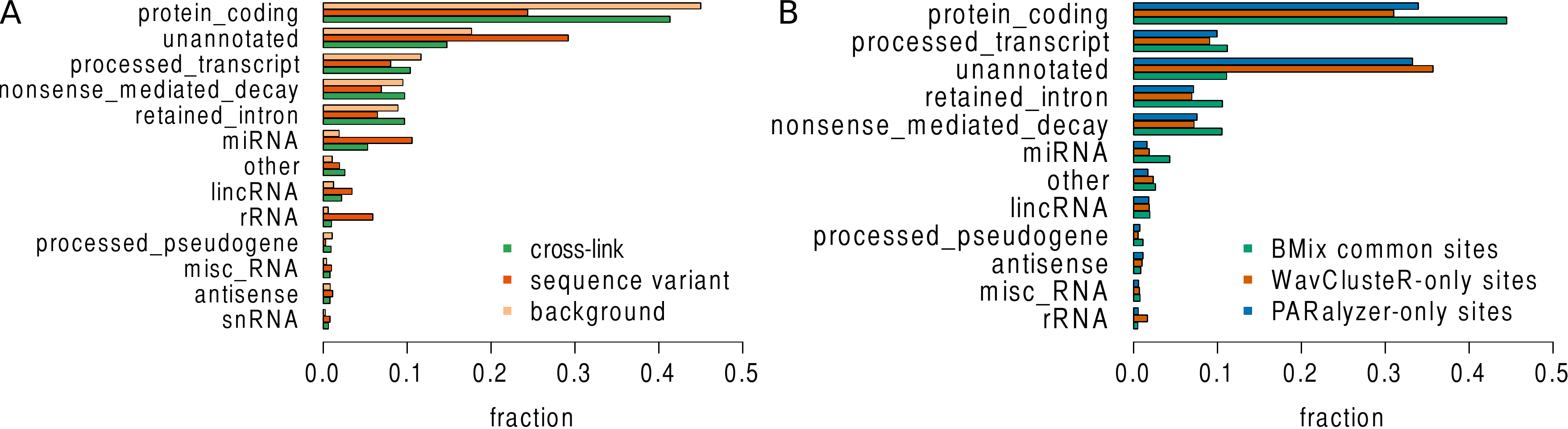}}
	\caption{ \textbf{(A)} Proportion of BMix-classified loci within each Ensembl gene type retrieved using the UCSC table browser~\citep{Karolchik2004} for replicate A of the AGO2 PAR-CLIP dataset~\citep{Kishore2011}. \textbf{(B)} Annotation according to the Ensembl gene types for binding sites commonly identified by BMix and the other three methods, as well as for the additional sites reported by PARalyzer and WavClusteR. The proportion of binding sites within each gene type is displayed. All the Ensembl types which contained less than 0.1\% sites were grouped under the name~\emph{"other"} and all the sites which did not fall within any annotation were marked as~\emph{"unannotated"}.}
	\label{fig_annotation}
\end{figure*}

\begin{figure*}[t!]
	\centering
	\includegraphics[width=0.9\textwidth]{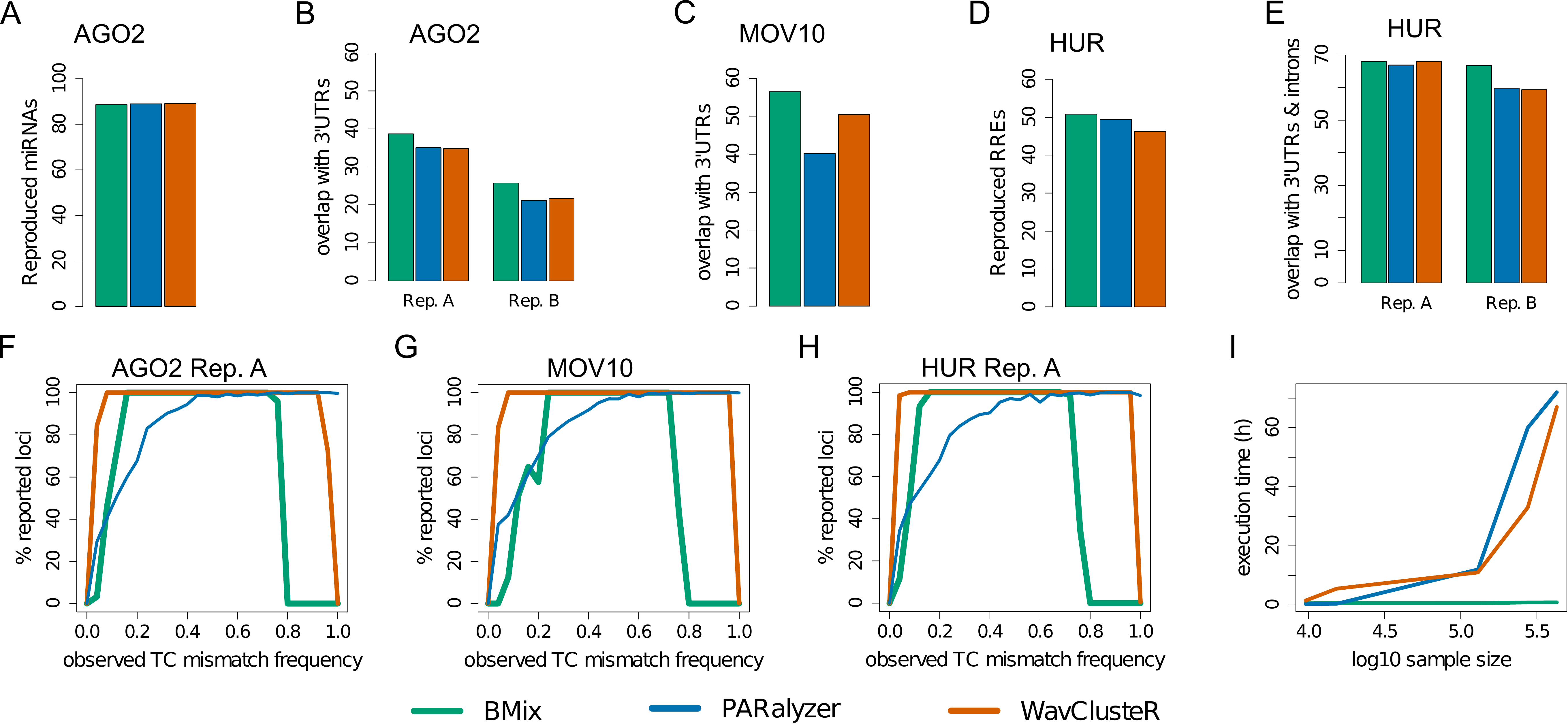}
	\caption{Summary of results from applying BMix, PARalyzer and WavClusteR on human datasets for proteins AGO2, MOV10 and HUR. \textbf{(A)} Reproducibility of the methods in terms of percentage of commonly reported miRNAs between two AGO2 PAR-CLIP replicates. \textbf{(B)} Percentage of binding sites reported in 3'UTRs for the AGO2 datasets. \textbf{(C)} Percentage of binding sites reported in 3'UTRs for the MOV10 dataset. \textbf{(D)} Reproducibility of the methods in terms of commonly reported~RREs between two HUR PAR-CLIP replicates. \textbf{(E)} Percentage of binding sites reported in 3'UTRs and introns for the HUR datasets. \textbf{(F-H)} Fraction of reported loci with a particular T-to-C substitution frequency out of the total number of observed loci with that substitution frequency. \textbf{(I)} Execution time for the three methods as function of log10 of the number of reported binding sites.}
	\label{all_results}
\end{figure*}

\noindent\textbf{Identification of AGO2 binding sites}\\
\begin{sloppypar}
Two replicate PAR-CLIP datasets were tested for the AGO2 protein. Since AGO2 is one of the proteins involved in RISC, its affinity to miRNAs and 3'UTRs was expected to be elevated~\citep{Hendrickson2008}.
\end{sloppypar}
\indent BMix identified 15,317 binding sites for replicate A, and 9,615 binding sites for replicate B of the AGO2 dataset. We annotated the three classes of loci reported by BMix (background, sequence variant and cross-link) according to the Ensembl gene types retrieved using the UCSC Table Browser~\citep{Karolchik2004}. A higher proportion of background and sequence variant loci overlapped with ribosomal RNA (rRNA) and unannotated regions than the cross-link loci which mostly covered protein-coding regions (Figure~\ref{fig_annotation}A). Both PARalyzer and WavClusteR reported around 4000 more binding sites than BMix (Supplementary Table 2). Annotation of the binding sites according to the same Ensembl types showed that a large proportion of these additional sites covered significantly more rRNA and unannotated regions and less protein coding regions compared to the common sites (Figure~\ref{fig_annotation}B). To assess the reproducibility of the three methods, each method was applied on each AGO2 replicate dataset independently, and the number of common miRNAs found within the binding sites between replicates was reported. All the three methods yielded a similar high percentage of reproducible miRNAs: 88.6\% for BMix, 88.98\% for PARalyzer and 89.1\% for WavClusteR (Figure~\ref{all_results}A and Supplementary Table 2). \\
\indent For replicate A, over 95.37\% of the binding sites found with BMix overlapped with the ones found by PARalyzer, and 99.99\% overlapped with WavClusteR. Similarly, for replicate B, 96.65\% of the sites reported by BMix overlapped with the PARalyzer binding sites, and 99.99\% with the WavClusteR sites (Supplementary Table 2). BMix reported on average 4\% more of its binding sites within 3'UTRs than the other two methods (Figure~\ref{all_results}B and Supplementary Table 2). Over 70\% of the binding sites reported by BMix were less than 30 nucleotides long (Supplementary Figure 8A), in concordance with the expected small length of miRNA targets. Furthermore, given that it has been previously reported that some targeted 3'UTRs may have high miRNA target-site abundance~\citep{Garcia2011}, we investigated the number of identified binding sites in each 3'UTR. We found that over 85\% of the covered 3'UTRs contained only one binding site, while a small proportion had several binding sites (Supplementary Figure 8B). 

\vspace{0.2cm}
\noindent\textbf{Identification of MOV10 binding sites} \\
Next, we applied the three methods on a dataset for the MOV10 protein, also expected to preferentially bind in 3'UTRs~\citep{cem}. In absence of replicates, we could not assess the reproducibility of the methods on this dataset. However, as for the previous dataset, BMix found a higher percentage of its binding sites (56.39\%) in 3'UTRs than PARalyzer and WavClusteR which obtained an overlap of 40.14\% and 50.4\%, respectively (Figure~\ref{all_results}C and Supplementary Table 3). Furthermore, more than 96\% of the binding sites reported by BMix overlapped with the ones found by the two other methods. 

\vspace{0.2cm}
\noindent\textbf{Identification of HUR binding sites} \\
Finally, we applied a similar evaluation scheme on two replicate PAR-CLIP datasets for the HUR protein, a well-characterized RBP involved in maintaining mRNA stability and regulating gene expression~\citep{peng1988}. It has been shown that this protein preferentially binds AU-rich regions in 3'UTRs of messenger RNAs, as well as intronic regions~\citep{Lebedeva2011340}. Therefore, we quantified the amount of binding sites found by each method within these genomic features for each replicate independently, as well as their enrichment for HUR-specific RNA recognition elements~(RREs) described in~\citep{Ma1996}. \\
\indent For both replicates, the binding sites found by BMix overlapped over 90\% with the binding sites reported by the other methods (Supplementary Table 4). To assess the reproducibility of the methods, we evaluated the percentage of RRE-enriched sites common between the two replicates. Even though it reported less binding sites than the other methods, BMix reached a higher reproducibility of 50.8\% compared to 49.45\% obtained with PARalyzer and 46.27\% reported by WavClusteR (Figure~\ref{all_results}D and Supplementary Table 4). For both replicates, BMix had also a superior percentage of RRE-enriched sites than the other two methods (Supplementary Table 4).\\ 
\indent BMix, PARalyzer and WavClusteR were applied on each replicate HUR dataset independently. For both replicates, BMix reported more binding sites within 3'UTRs and intronic regions compared to the other two methods (Figure~\ref{all_results}E). Precisely, 68.1\% of the sites reported by BMix for replicate A overlapped with 3'UTRs and intronic regions, while PARalyzer and WavClusteR attained 66.95\% and 68.06\%, respectively (Supplementary Table 4). For replicate B, despite the lower number of reported sites, BMix had a superior overlap with 3'UTRs, namely 66.83\%, compared to 59.83\% and 59.37\% obtained by PARalyzer and WavClusteR, respectively (Supplementary Table 4).\\

\noindent For all the datasets, the additional cross-link loci reported by PARalyzer and WavClusteR compared to BMix had either a low substitution frequency and low coverage, or a high substitution frequency. To illustrate this, we calculated, for each method, the fraction of reported loci with a particular T-to-C substitution frequency out of the total number of observed loci with that substitution frequency. PARalyzer reported all the high-frequency altered loci with no exception, while WavClusteR learned a more lenient threshold than BMix for the low and high substitution frequency regions (Figures~\ref{all_results}F-H and Supplementary Figure S5). Furthermore, the additional binding sites reported by the two other methods were more prevalent in rRNA and unannotated regions compared to the BMix sites (Figure~\ref{fig_annotation}B and Supplementary Figure S7). A large proportion of the binding sites reported by BMix were less than 30 nucleotides long, and several covered 3'UTRs contained more than one binding site (Supplementary Figures S8-S12).\\ 
\indent We analyzed the impact of the alignment strategy on the results of our method by testing BMix on Bowtie-aligned AGO2 PAR-CLIP data~\citep{Kishore2011}, allowing for one, two and three mismatches. The number of BMix-reported binding sites increased to over 40\%, from 15,317 sites with one mismatch to 21,607 and 25,289 sites with two and three mismatches, respectively. Over 50\% of the identified binding sites with two or three mismatches were also detected with one mismatch (Supplementary Table 5). More than 85\% of the binding sites identified with BMix using one mismatch were also found by using two or three mismatches. Nevertheless, the configuration of the three types of loci classified with BMix changed as the number of mismatches was increased. Precisely, the fraction of reported cross-linked loci decreased, whereas the other two classes gained in size as the number of allowed mismatches increased (Supplementary Figure S13A). On the other hand, PARalyzer doubled the number of reported binding sites when more than one mismatch was allowed, reporting 19,248 sites for one mismatch, 40,522 sites for two mismatches, and 51,029 sites for three mismatches. WavClusteR functions specifically for alignments with one allowed mismatch, therefore testing for more mismatches was not possible. We performed the same analysis with BMix by choosing TopHat~\citep{Trapnell2009} as aligner (Supplementary Figure S13B) on the same dataset, although PAR-CLIP reads are generally too short to be efficiently used by a splice-aware aligner. Similar results were obtained with TopHat as with Bowtie, especially when one or two mismatches were allowed (Supplementary Tables 5 and 6, and Supplementary Figure S13). \\
\indent In terms of execution time, on one core of a Linux machine with a clock rate of 2.3GHz, BMix proved to be considerably faster than the other two methods, running on average in less than 40min on all the datasets. On the contrary, the execution time of both PARalyzer and WavClusteR on the same machine increased with the sample size from several hours to multiple days for the HUR datasets (Figure~\ref{all_results}I).\\

\section{Discussion}
In the presented work, we have proposed BMix, a new probabilistic model for identifying high-confidence RNA-protein interaction sites from PAR-CLIP data. BMix uses a constrained semi-supervised three-component binomial mixture model to describe the T-to-C substitutions observed at genomic loci in three categories: low-frequency errors due to sequencing noise, true cross-link sites, or high-frequency sequence variants caused by SNPs or contamination. Therefore, our model brings the novelty of accounting for both low and high-frequency erroneous T-to-C alterations in PAR-CLIP data. We validated and demonstrated the superior performance of BMix compared to the methods WavClusteR v2.0.0 and PARalyzer v1.1 both on synthetic and real data.\\
\indent Most of the current PAR-CLIP analysis methods focus on filtering low-frequency altered loci and consider the high-frequency substitutions as reliable indicators of cross-linking. We have observed this behavior also within our study on real data, where PARalyzer has selected all the high-frequency altered loci within its reported binding sites (Figures~\ref{all_results}F-H and Supplementary Figure S5). However, methods like PARalyzer ignore the possibility that high-frequency alterations could have been caused by other factors such as contamination or single nucleotide variants. By comparing PAR-CLIP and matched control RNA-Seq data from the same experiment, the prevalence of highly altered loci was clearly observed also for non T-to-C substitutions in published data (Figure~\ref{AC}), which motivates the need of identifying and discarding spurious highly altered loci from the analysis. So far, only the WavClusteR method accounts for high-frequency substitutions. However, because it uses relative substitution frequency values instead of actual read counts, the method looses performance especially in low and high substitution regions, as presented in real and synthetic data applications (Figures~\ref{accuracy},~\ref{all_results}F-H and Supplementary Figures S2 and S5). \\
\indent In an extensive simulation study using varying PAR-CLIP substitution probability $\mu$, we have observed that the accuracy of both BMix and WavClusteR attains a global maximum at $\mu=0.6$ and $\mu=0.7$, respectively and then decreases. In other words, a perfect 100\% PAR-CLIP T-to-C substitution efficiency would not improve the result; on the contrary it would make difficult to differentiate between induced substitutions and high-frequency errors.\\
\indent The Ensembl annotation of the three classes of loci reported by BMix showed that our model captures more rRNA and unannotated RNA within its background and sequence variant mixture components, while the reported cross-link loci mainly cover protein-coding regions and miRNAs. The PARalyzer approach, based on selecting high-frequency mutations as high-confidence binding sites is therefore at risk of reporting a large amount of spurious alterations as cross-link loci. WavClusteR is exposed to the same risk by using relative frequencies instead of substitutions counts, thus disregarding uncertainty in the low coverage regions. As a result, BMix identified less binding sites than the other two methods, at the same time keeping high the proportion of binding sites overlapping with features of interest. The additional binding sites detected by the other two methods overlapped more with rRNA and unannotated regions than the sites commonly identified with BMix.\\
\indent BMix, as well as the other PAR-CLIP analysis methods, relies on the alignment of sequencing reads to a reference genome. In this work, we have chosen the standard alignment strategy employed by PARalyzer and WavClusteR, aligning the sequencing reads with Bowtie and allowing for one mismatch. However, a strict alignment strategy would potentially discard reads with viable cross-link T-to-C alterations. Due to explicit modeling of noise and contamination, our model is less sensitive to the choice of alignment and is able to control the false positive rate even for more lenient alignment parameters. As a result, the user of BMix can pick the alignment procedure which best suits the data without having to maintain a too strict control on the alignment parameters. This procedure depends on multiple aspects such as, for example, the quality of the sequencing, the length of the reads, or the binding protein~\citep{Konig2012}. Ultimately, a systematic comparison of the results obtained from different alignment strategies can be utilized for a better quantification of the binding sites.\\
\indent A limitation of our method consists in the difficulty of sorting out T-to-C substitutions of moderate frequency which have not been introduced by PAR-CLIP. These can occur following diverse molecular processes such as, for example, RNA editing by ADAR enzyme~\citep{Samuel2011, B1}, following contamination, or as heterozygous SNPs. In this case, any analysis relying entirely on coverage and substitution counts does not have sufficient statistical power to discard these false loci, and our method can report false positives originating from these loci. Nevertheless, these substitutions would also appear during a control experiment. A control PAR-CLIP experiment that would include the same steps as in the normal PAR-CLIP protocol, except for inducing T-to-C alterations, could be helpful. Therefore, a potential solution to this limitation would be to use a control PAR-CLIP or RNA-seq experiment. The control experiments would reveal these substitutions and one can afterwards subtract them from the PAR-CLIP cross-link T-to-C alterations identified with BMix.\\
\indent Due to the reduction of sequencing costs, a high increase in sequencing-based experiments such as PAR-CLIP is expected in the near future. BMix provides a rigorous probabilistic method which is significantly faster and more accurate than the current state-of-the art methods for detecting RNA-protein interaction sites in PAR-CLIP data.


\section*{Acknowledgement}
The authors thank Federico Comoglio and Cem Sievers for valuable feedback and discussions, as well as their support with WavClusteR.

\bibliographystyle{natbib}
\normalsize
\bibliography{references}

\newpage
\setcounter{figure}{0}


\graphicspath{{images//}}

\begin{center}
	\Large
	Supplementary materials for: \\ Probabilistic modeling of occurring substitutions in PAR-CLIP data \\ \normalsize Monica Golumbeanu, Pejman Mohammadi, Niko Beerenwinkel
\end{center}

\makeatletter 
\renewcommand{\thefigure}{S\@arabic\c@figure}
\makeatother

\subsection*{Supplementary Figures}

\begin{figure}[h!]
	\centering
	\includegraphics[scale=0.6]{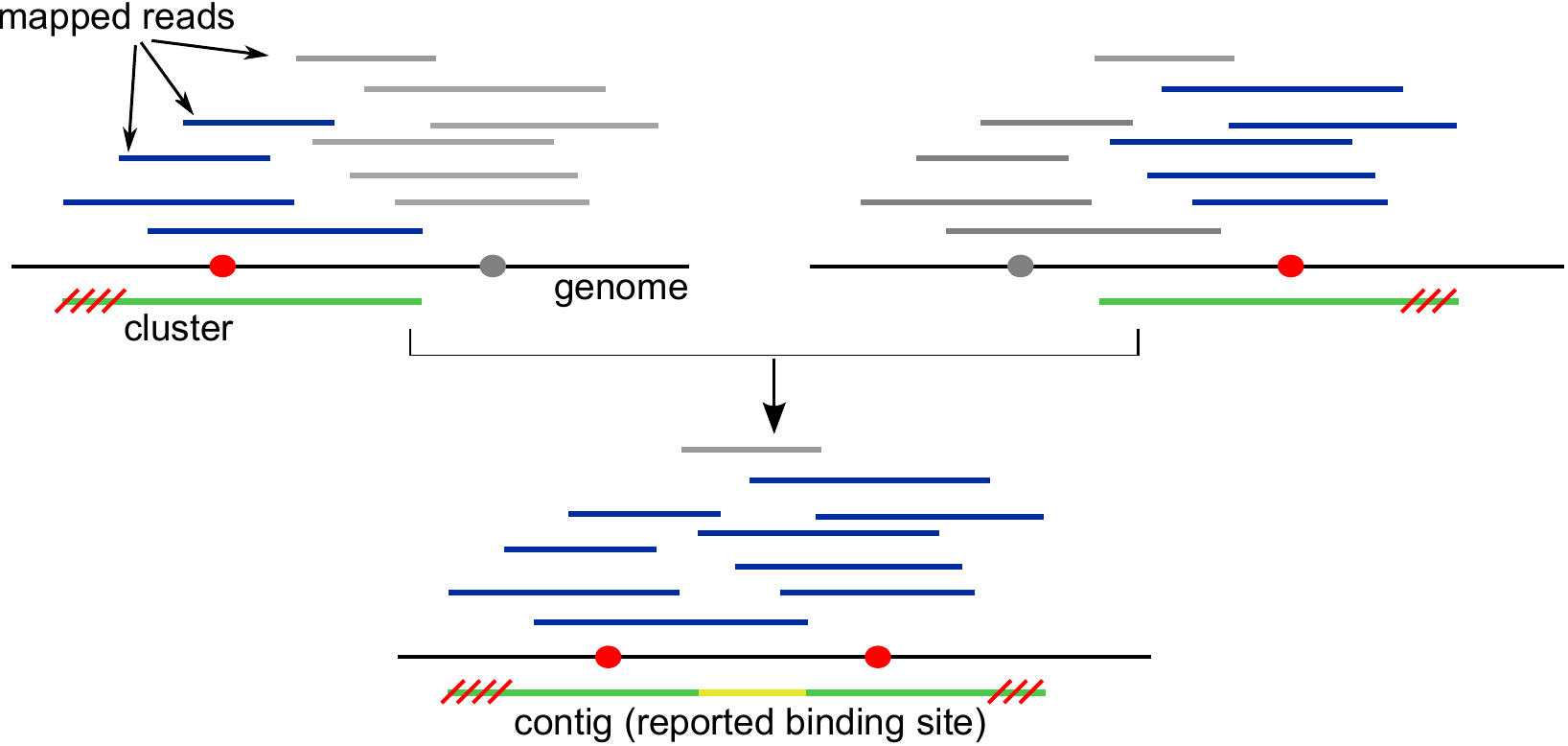}
	\caption{Construction of a binding site starting from high-confidence cross-link T loci (red circles). First, the reads spanning these loci (blue segments) are grouped into clusters. Low coverage margins are trimmed off. Overlapping clusters are merged into contigs and reported as ~RNA-RBP binding sites.}
	\label{S1}
\end{figure}

\begin{figure}[h!]
	\centering
	\includegraphics[scale=0.35]{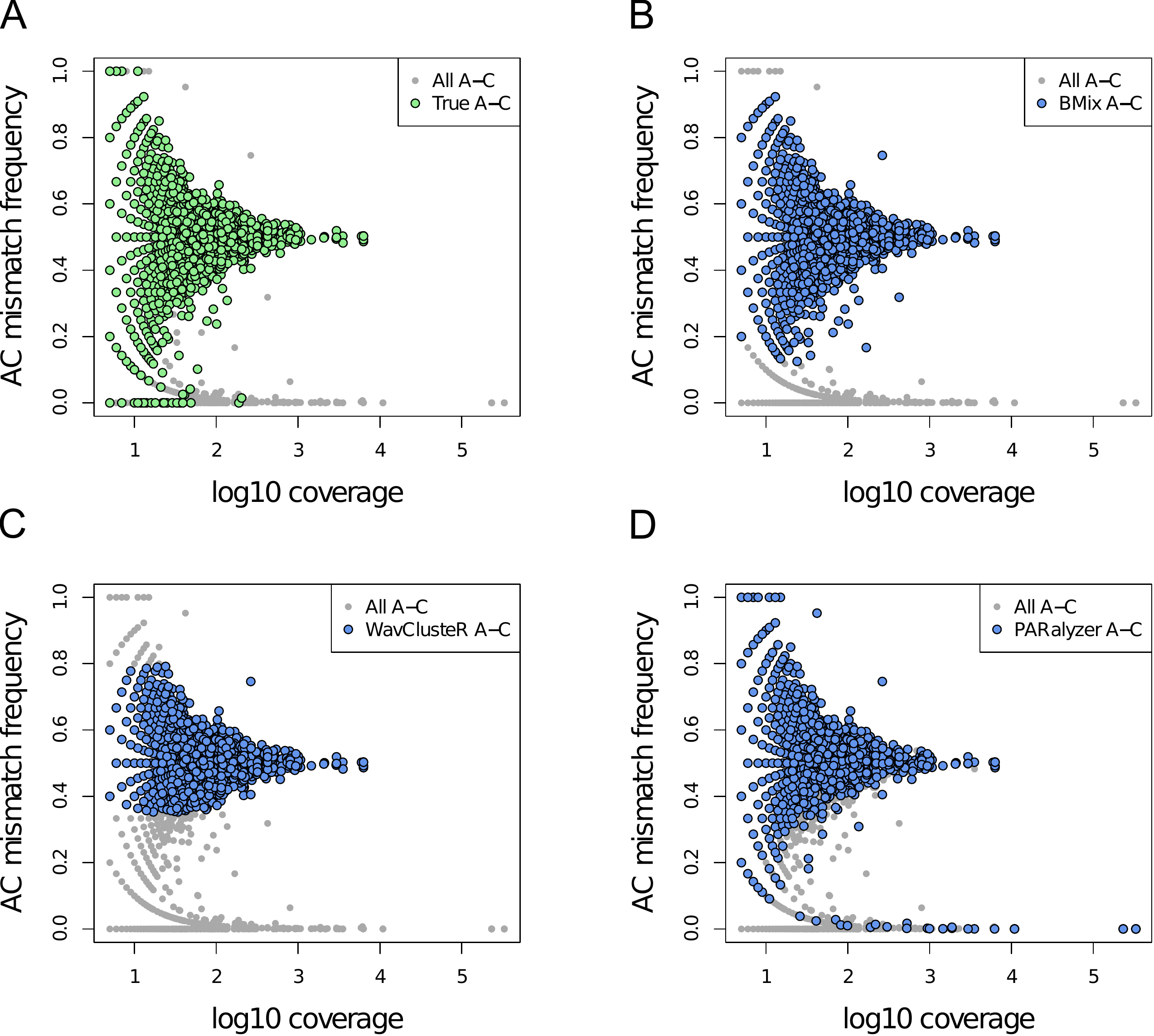}
	\caption{BMix, PARalyzer and WavClusteR outcomes on a simulated dataset with incorporated A-to-C alterations with $\mu=0.5$ probability. In all figures, for each genomic locus, the observed A-to-C substitution frequency is represented as function of log10 of the coverage. ~\textbf{(A)} True A-to-C induced substitutions (green) within the simulated data (grey).  ~\textbf{(B)} A-to-C substitutions detected by BMix (blue). ~\textbf{(C)} A-to-C substitutions identified by WavClusteR (blue). ~\textbf{(D)} A-to-C substitutions reported by PARalyzer (blue).}
	\label{S2}
\end{figure}

\newpage

\begin{figure}[h!]
	\centering
	\includegraphics[scale=0.5]{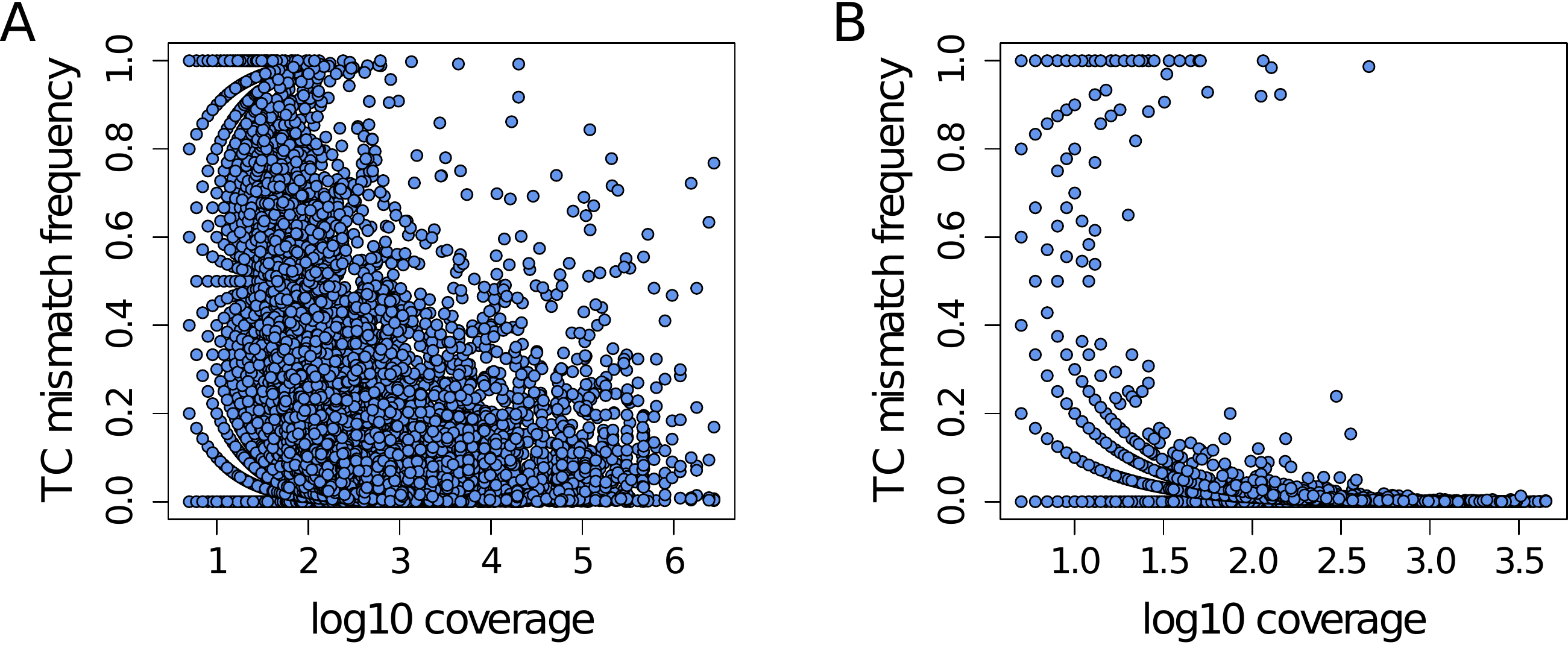}
	\caption{Genome-wide T-to-C observed mismatch frequency as a function of log10 of coverage in a PAR-CLIP dataset~\textbf{(A)} compared to the matched control RNA-Seq dataset~\textbf{(B)}.}
	\label{S3}
\end{figure}

\begin{figure}[h!]
	\centering
	\includegraphics[scale=0.5]{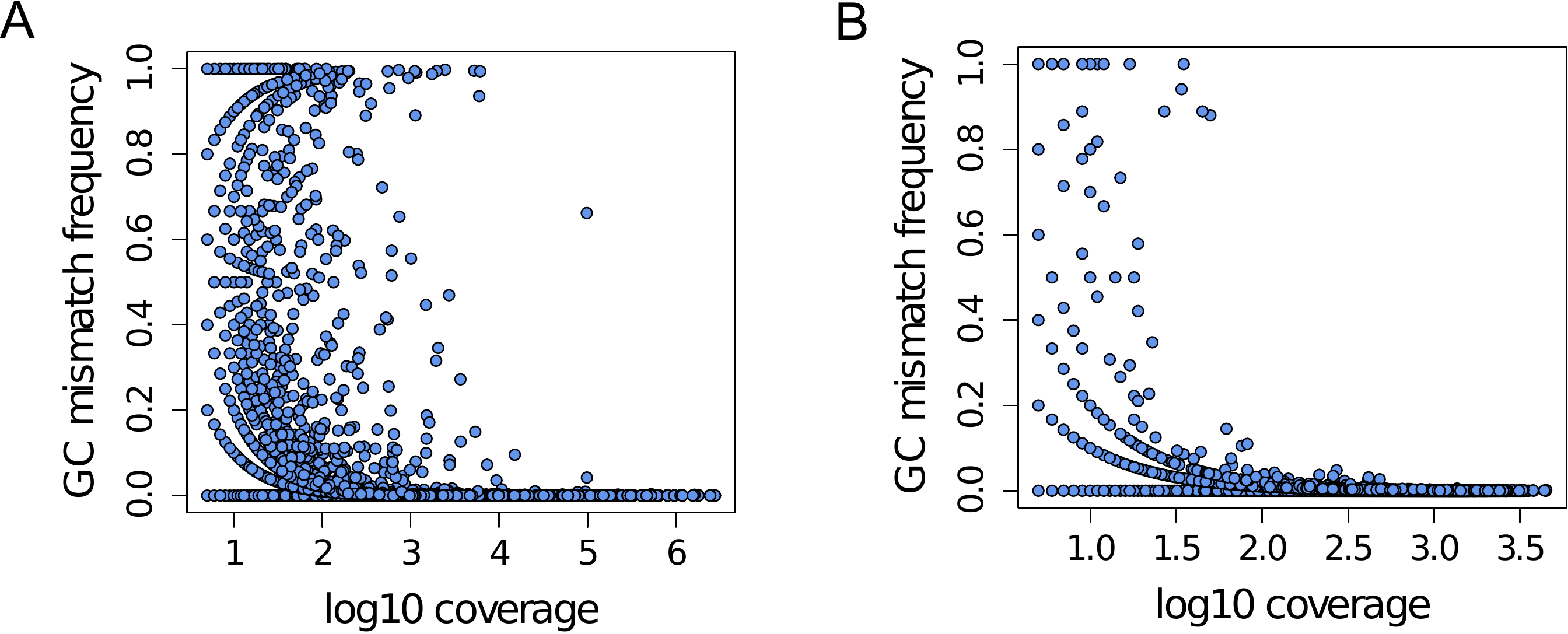}
	\caption{Genome-wide G-to-C observed mismatch frequency as a function of log10 of coverage in a PAR-CLIP dataset~\textbf{(A)} compared to the matched control RNA-Seq dataset~\textbf{(B)}.}
	\label{S4}
\end{figure}

\begin{figure}[h!]
	\centering
	\includegraphics[scale=0.4]{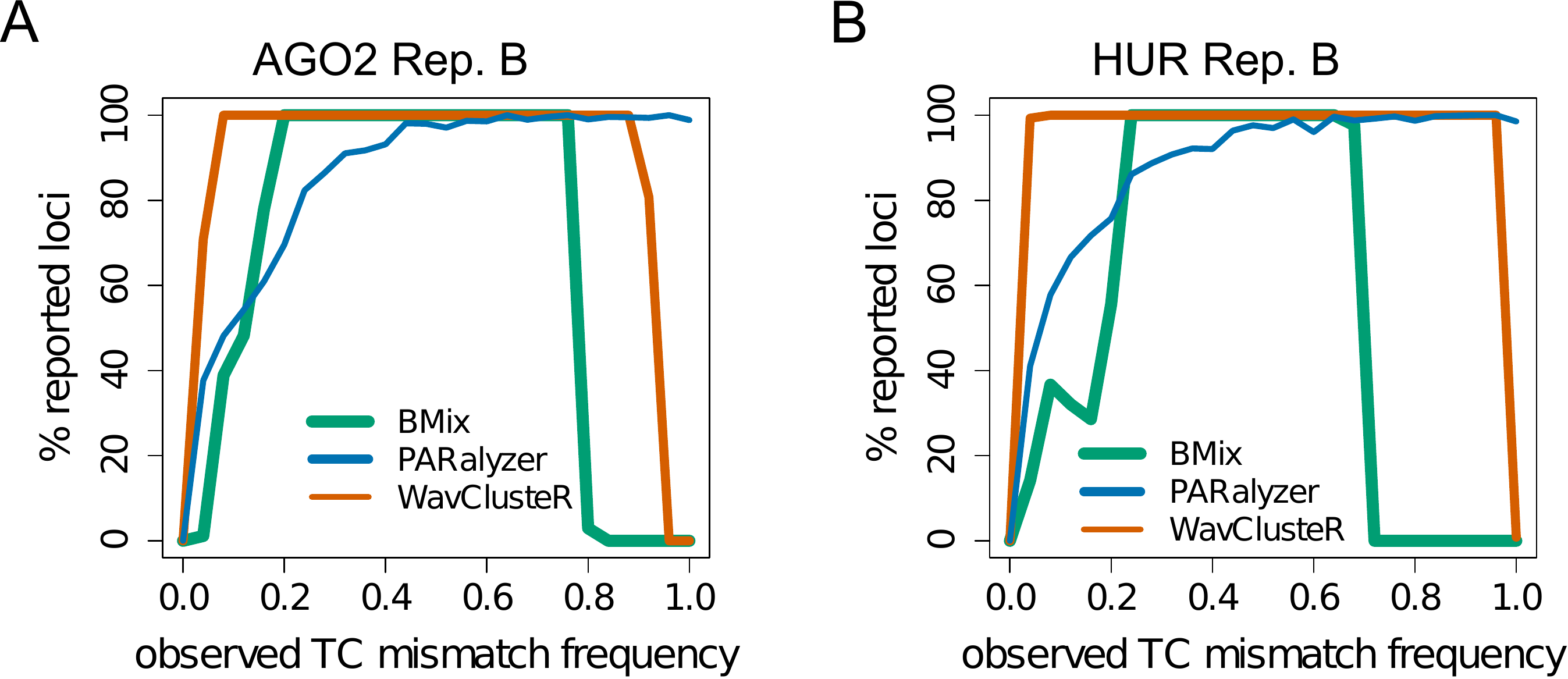}
	\caption{Fraction of reported loci with a particular T-to-C substitution frequency out of the total number of observed loci with that substitution frequency for two PAR-CLIP datasets from proteins AGO2 and HUR.}
	\label{S5}
\end{figure}

\newpage

\begin{figure}[h!]
	\centering
	\includegraphics[width = 0.8\textwidth]{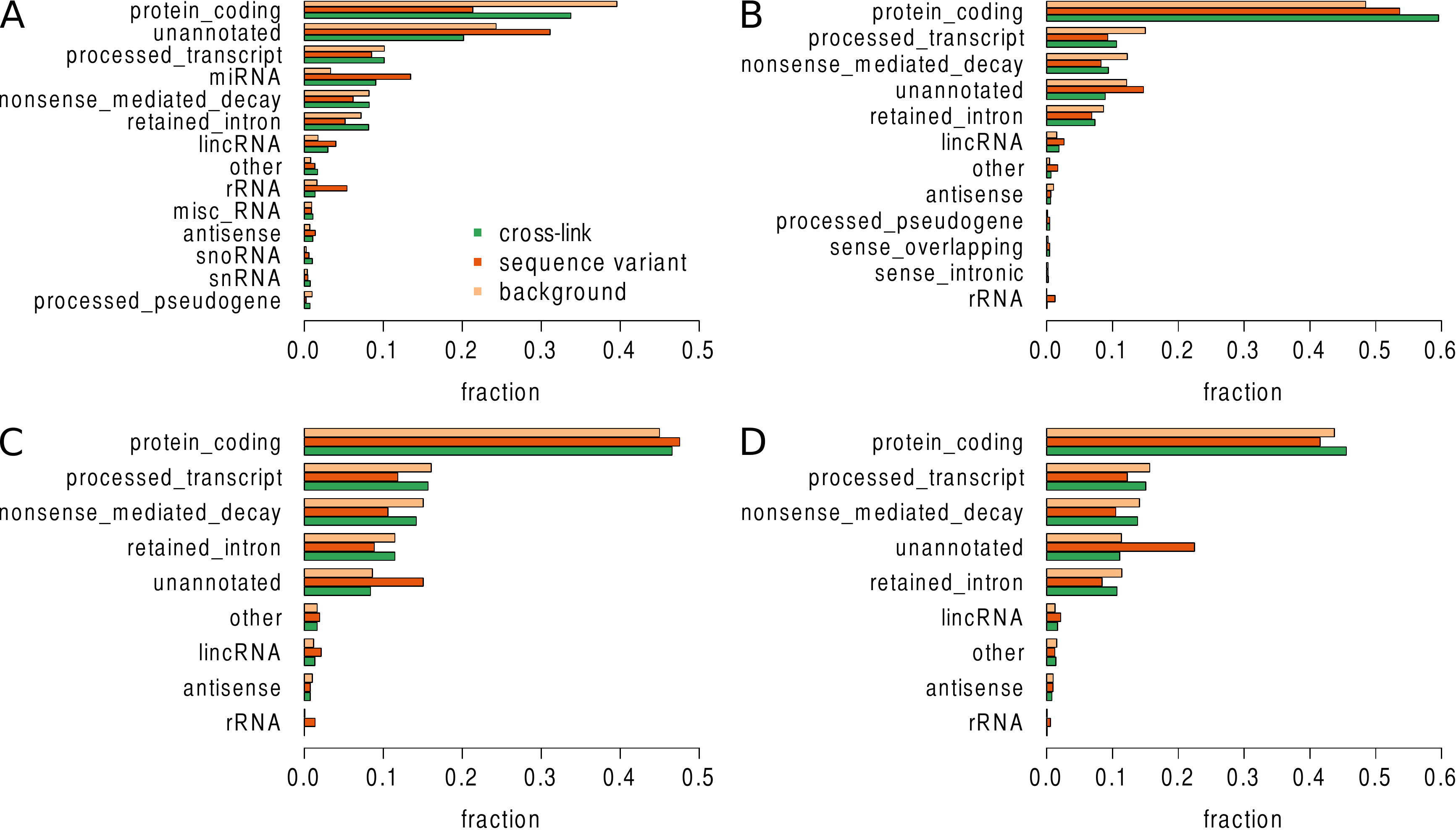}
	\caption{Proportion of BMix-classified loci within each Ensembl gene type retrieved using the UCSC Table Browser for the replicate B of the AGO2 dataset \textbf{(A)}, MOV10 dataset \textbf{(B)}, replicate A of the HUR dataset \textbf{(C)} and replicate B of the HUR dataset \textbf{(D)}. All the Ensembl types which contained less than 0.1\% loci were grouped under the name~\emph{"other"} and all the loci which did not fall within any annotation were marked as~\emph{"unannotated"}.}
	\label{S6}
\end{figure}

\begin{figure}[h!]
	\centering
	\includegraphics[width = 0.8\textwidth]{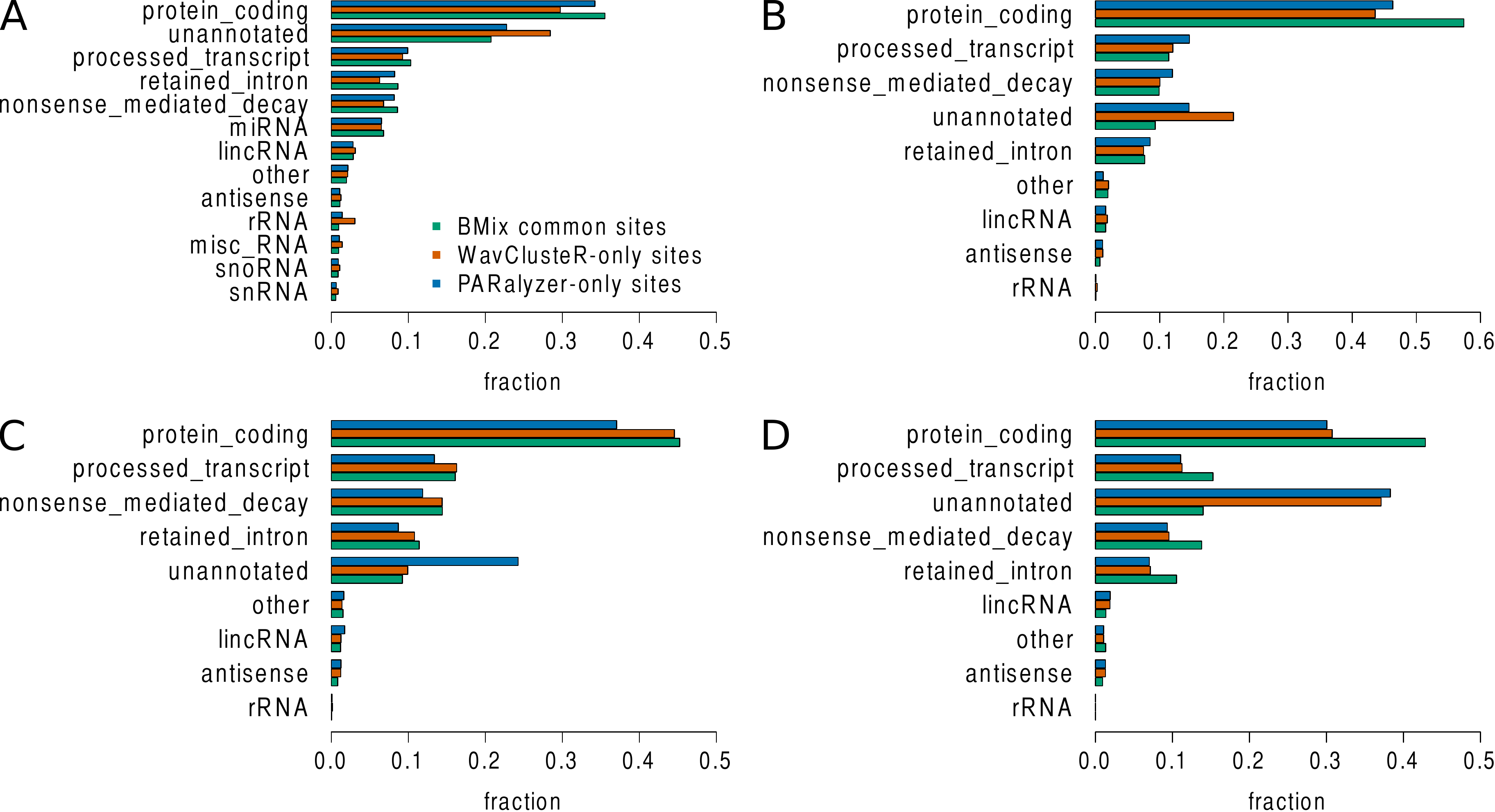}
	\caption{Annotation according to the Ensembl gene types for binding sites commonly identified by BMix and the other two methods, as well as for the additional sites reported by PARalyzer and WavClusteR for the replicate B of the AGO2 dataset \textbf{(A)}, MOV10 dataset \textbf{(B)}, replicate A of the HUR dataset \textbf{(C)} and replicate B of the HUR dataset \textbf{(D)}. The proportion of binding sites within each gene type is displayed. All the Ensembl types which contained less than 0.1\% sites were grouped under the name~\emph{"other"} and all the sites which did not fall within any annotation were marked as~\emph{"unannotated"}.}
	\label{S7}
\end{figure}
\newpage

\begin{figure}[h!]
	\centering
	\includegraphics[width = 0.7\textwidth]{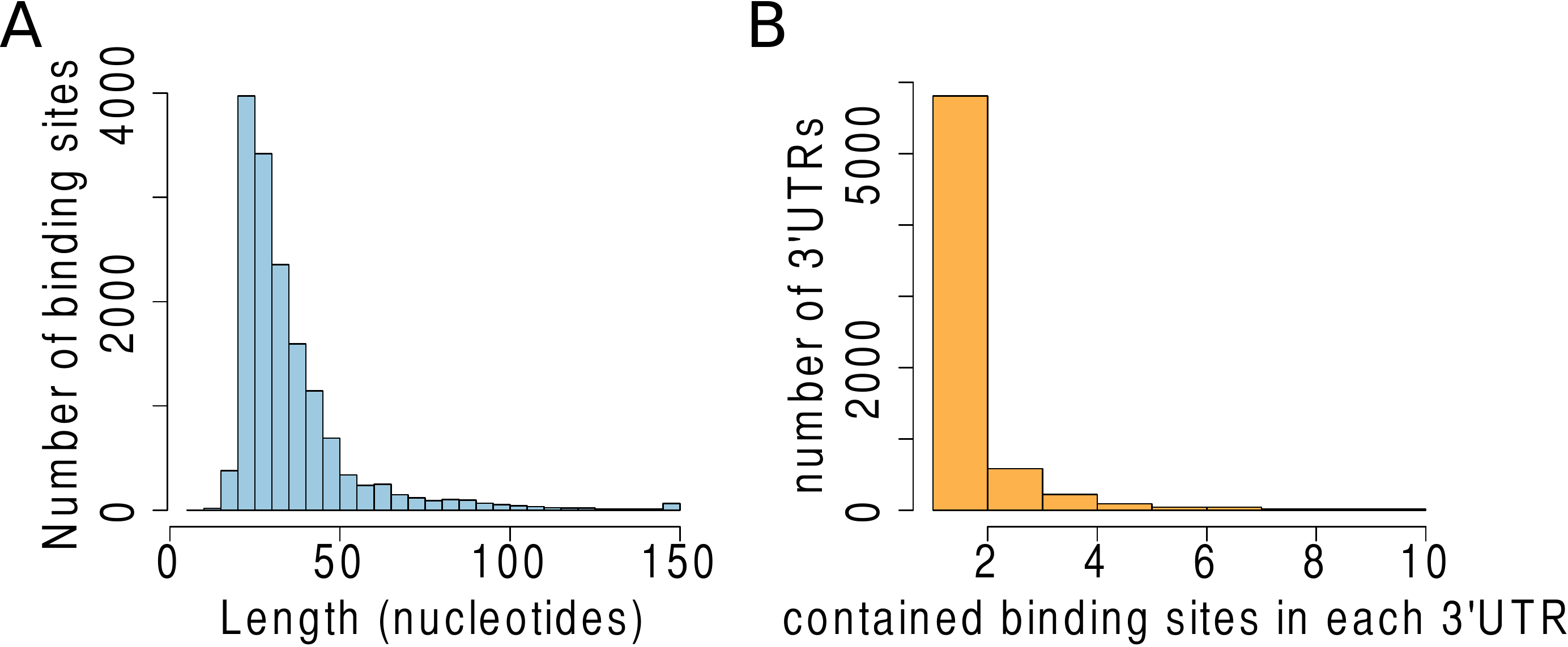}
	\caption{\textbf{(A)} Histogram of the length of the binding sites reported by BMix for replicate A of the AGO2 dataset. \textbf{(B)} Histogram of the number of binding sites reported by BMix in each 3'UTR for replicate A of the AGO2 dataset.}
	\label{S8}
\end{figure}

\begin{figure}[h!]
	\centering
	\includegraphics[width = 0.7\textwidth]{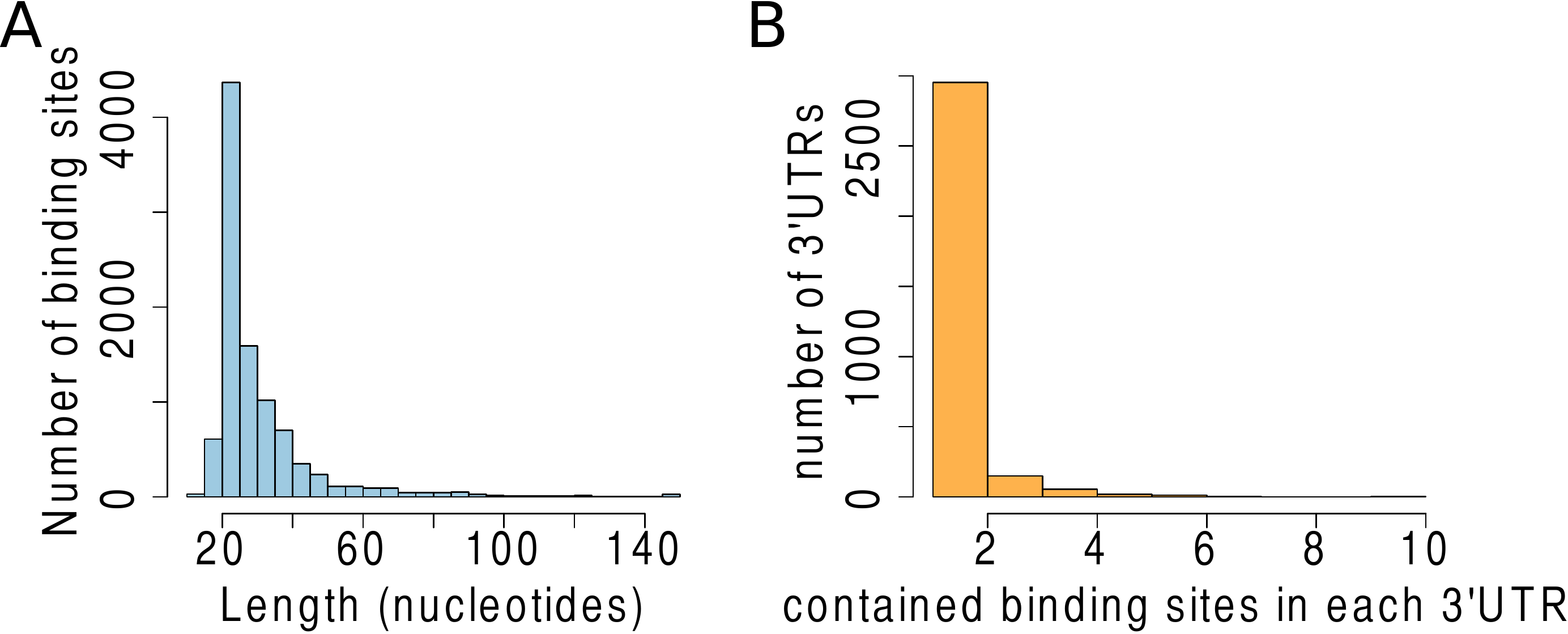}
	\caption{\textbf{(A)} Histogram of the length of the binding sites reported by BMix for replicate B of the AGO2 dataset. \textbf{(B)} Histogram of the number of binding sites reported by BMix in each 3'UTR for replicate B of the AGO2 dataset.}
	\label{S8}
\end{figure}

\begin{figure}[h!]
	\centering
	\includegraphics[width = 0.7\textwidth]{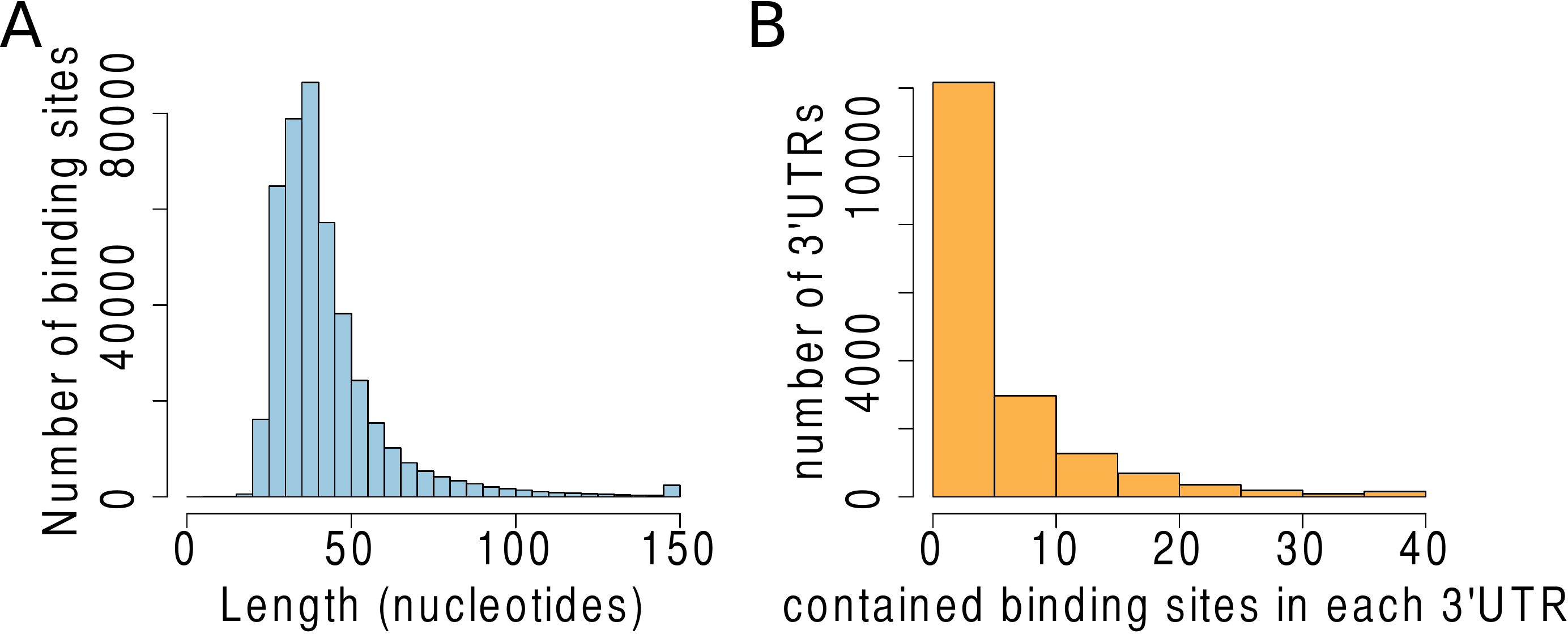}
	\caption{\textbf{(A)} Histogram of the length of the binding sites reported by BMix for replicate A of the HUR dataset. \textbf{(B)} Histogram of the number of binding sites reported by BMix in each 3'UTR for replicate A of the HUR dataset.}
	\label{S9}
\end{figure}

\newpage

\begin{figure}[h!]
	\centering
	\includegraphics[width = 0.7\textwidth]{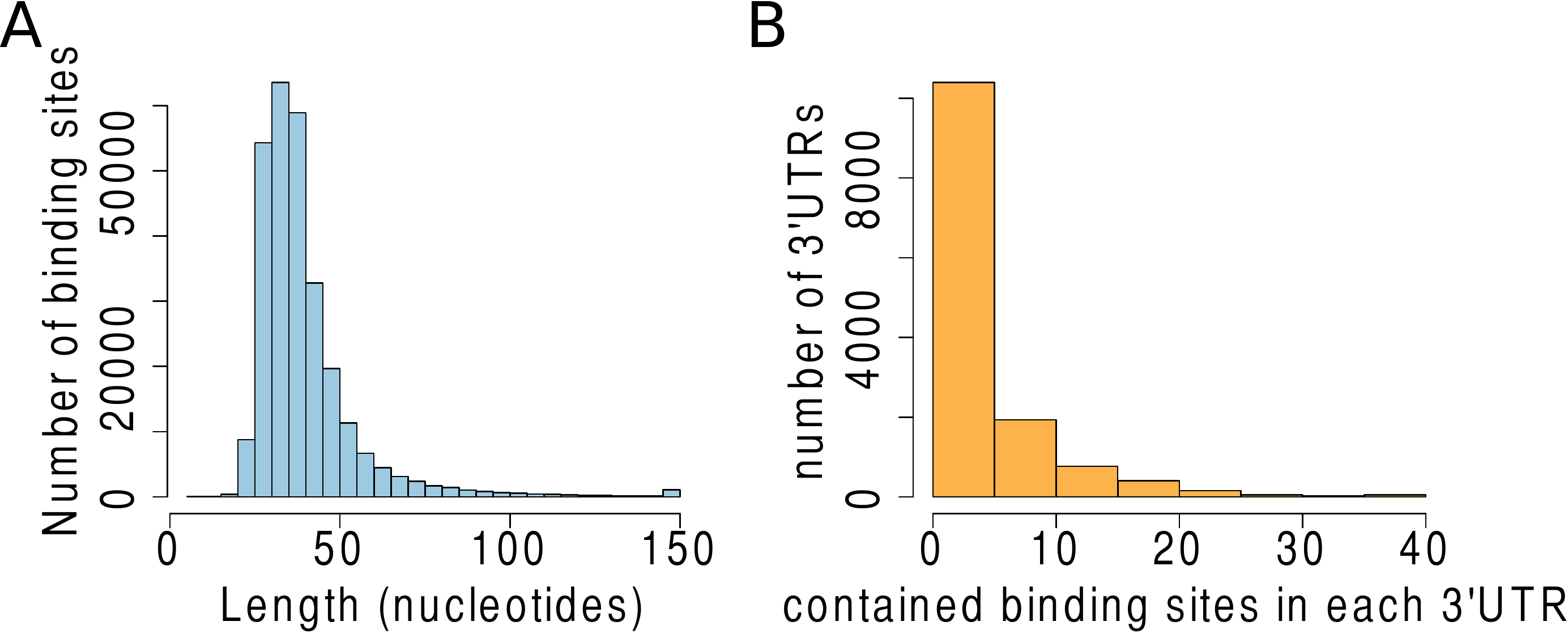}
	\caption{\textbf{(A)} Histogram of the length of the binding sites reported by BMix for replicate B of the HUR dataset. \textbf{(B)} Histogram of the number of binding sites reported by BMix in each 3'UTR for replicate A of the HUR dataset.}
	\label{S10}
\end{figure}

\begin{figure}[h!]
	\centering
	\includegraphics[width = 0.7\textwidth]{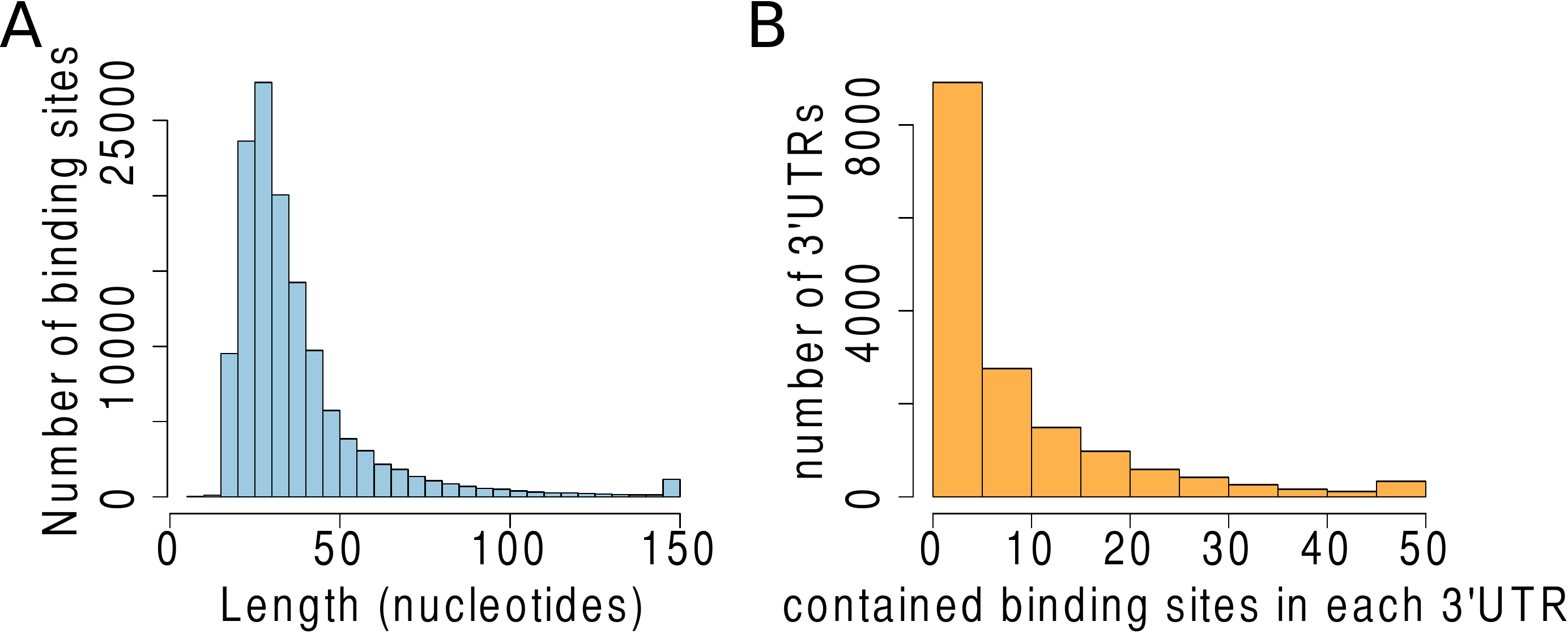}
	\caption{\textbf{(A)} Histogram of the length of the binding sites reported by BMix for the MOV10 dataset. \textbf{(B)} Histogram of the number of binding sites reported by BMix in each 3'UTR for the MOV10 dataset.}
	\label{S11}
\end{figure}

\begin{figure}[h!]
	\centering
	\includegraphics[width = 0.7\textwidth]{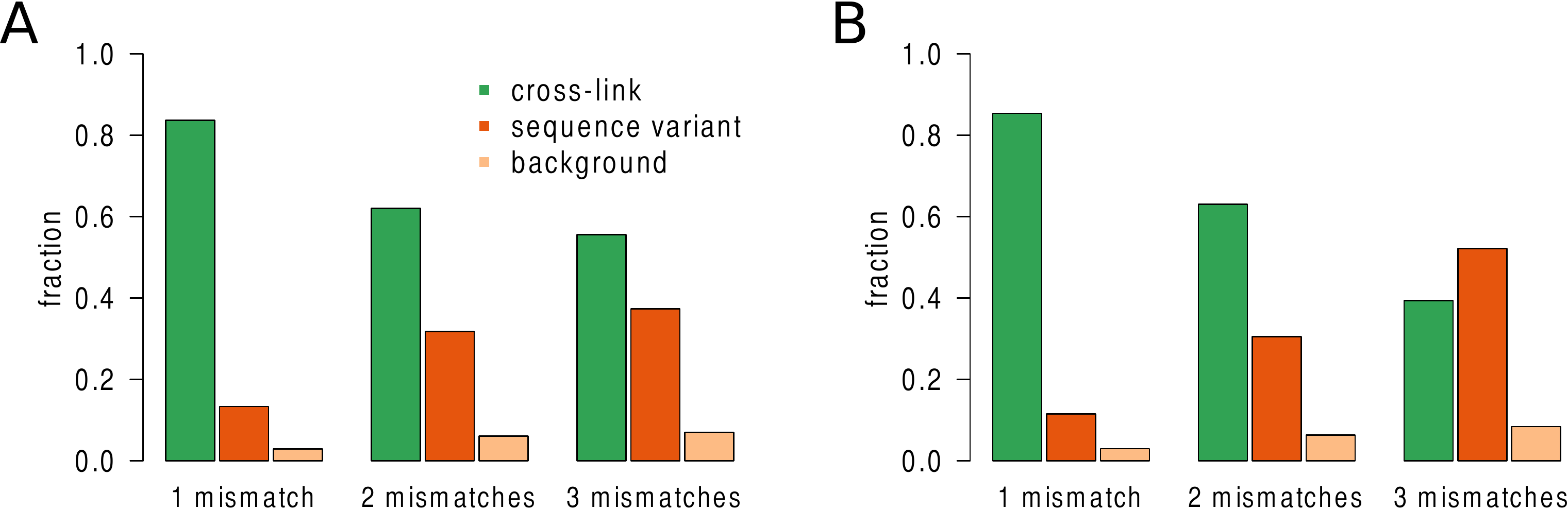}
	\caption{\textbf{(A)} Proportion of each BMix mixture component out of the total number of classified loci when Bowtie is used to align the PAR-CLIP reads with one, two or three allowed mismatches. \textbf{(B)} Proportion of each BMix mixture component out of the total number of classified loci when TopHat is used to align the PAR-CLIP reads with one, two or three allowed mismatches. }
	\label{S12}
\end{figure}
\newpage

\subsection*{Supplementary Tables}

\begin {table}[h!]
\begin{center}
	\begin{tabular}{ c  c  c  c  c  c  c  c }
		\hline
		dataset & replicate & strand & $\epsilon$ & $\gamma$ & $\theta$ & $\nu$ & $\psi$ \\ \hline
		AGO2 & A & + & 0.001 & 0.156 & 0.1564 & 0.0056 & 0.3462 \\
		AGO2 & A & - & 0.001 & 0.1508 & 0.1512 & 0.0061 & 0.3693 \\
		\hline
		AGO2 & B & + & 0.001 & 0.1695 & 0.1698 & 0.0063 & 0.2613 \\
		AGO2 & B & - & 0.001 & 0.1675 & 0.1678 & 0.0069 & 0.2641 \\
		\hline
		MOV10 & & + & 0.0022& 0.217 & 0.2173& 0.016 & 0.1531 \\
		MOV10 & & - & 0.0021& 0.216 & 0.2163 & 0.0167 & 0.1556 \\
		\hline
		HUR & A & + & 0.002 & 0.0973 & 0.1 & 0.001 & 0.5372 \\
		HUR & A & - & 0.0018 & 0.1018 & 0.103 & 0.001 & 0.5953 \\
		\hline
		HUR & B & + & 0.0028 & 0.0625 & 0.065 & 0.001 & 0.4474 \\
		HUR & B & - & 0.0027 & 0.0656 & 0.0676 & 0.0011 & 0.52 \\
		\hline
	\end{tabular}
	\caption {Estimated parameters (see Methods) of the BMix probabilistic model on real PAR-CLIP datasets for proteins AGO2, MOV10 and HUR.}
	\label{table_param}
\end{center}
\end {table}

\begin {table}[h!]
\begin{center}
	\begin{tabular}{ c  l  c  c  c }
		\hline
		& & BMix & PARalyzer & wavClusteR\\ \hline
		&No. sites & 15,317 & 19,248 & 19,312 \\ 
		&overlap with BMix & 100\% & 95.37\% & 99.99\% \\ 
		Replicate A&No. miRNA & 1024 & 1059 & 1055\\ 
		& in 3'UTRs & 38.67\% & 35.02 \% & 34.83\%\\ 
		& in 5'UTRs & 8\% & 7.2\% & 7.9\% \\
		&execution time & 40min & 23min & 5h30min\\ 
		\hline
		&No. sites & 9,615 & 13,381 & 13,030 \\ 
		&overlap with BMix & 100\% & 96.65\% & 99.99\% \\ 
		Replicate B&No. miRNA & 935 & 976 & 971\\ 
		&in 3'UTRs & 25.74\% & 21.15\% & 21.76\%\\ 
		&in 5'UTRs & 8.12\% & 6.84\% & 7.41\% \\
		&execution time & 35min & 18min & 1h30min\\ 
		\hline
		Replicate A vs. B&miRNA & 866 (88.6 \%) & 904 (88.98\%) & 901 (89.1\%)\\
		\hline
	\end{tabular}
	\caption {Result of running BMix, WavClusteR and PARalyzer on the AGO2 PAR-CLIP dataset with two replicates. The number of binding sites, overlap with BMix, their occurrence in 3'UTR and covered miRNAs are reported, as well as the execution time of each method. The overlap with BMix equals to the percentage of binding sites reported by BMix that overlap by at least one nucleotide with the binding sites of other methods. The last line presents an analysis of reproducibility of the three different methods by assessing the percentage of coomon miRNAs reported by each method in both replicates.}
	\label{tableAGO}
\end{center}
\end {table}
\newpage

\begin {table}[h!]
\begin{center}
	\begin{tabular}{  l  c  c  c }
		& BMix & PARalyzer & wavClusteR\\ \hline
		No. sites & 129,772 & 306,504 & 184,579 \\ 
		overlap with BMix & 100\% & 96.76\% & 99.99\% \\ 
		in 3'UTRs & 56.39\% & 40.14\% & 50.4\%\\ 
		execution time & 37min & 12h & 11h\\ 
	\end{tabular}
	\caption {Results of running BMix, WavClusteR and PARalyzer on the MOV10 PAR-CLIP dataset. The methods were assessed in terms of the amount of reported binding sites, their overlap with BMix, the presence of binding sites in 3'UTRs and execution time. The overlap with BMix equals to the percentage of binding sites reported by BMix that overlap by at least one nucleotide with the binding sites of other methods. }
	\label{tableMOV}
\end{center}
\end {table}

\begin {table}[h!]
\begin{center}
	\begin{tabular}{ c  l  c  c  c }
		\hline
		& & BMix & PARalyzer & wavClusteR\\ \hline
		&No. sites & 427,525 & 487,771 & 424,431 \\ 
		&overlap with BMix & 100\% & 91.67\% & 90.36\% \\ 
		Replicate A&in 3'UTRs or introns & 68.1\% & 66.95 \% & 68.06\% \\ 
		&RRE rich sites & 28.47\% & 18.77\% &19.97\% \\ 
		&execution time & 52min & 72h & 67h\\ 
		\hline
		&No. sites & 275,691 & 435,759 & 419,805 \\ 
		&overlap with BMix & 100\% & 94.44\% & 99.99\% \\ 
		Replicate B&in 3'UTRs or introns & 66.83\% & 59.83\% & 59.37\%\\ 
		&RRE rich sites & 42.56\% & 36.06\% & 41.03\%\\ 
		&execution time & 50min & 60h & 33h\\ 
		\hline
		Replicate A vs. B&No. RREs & 60,695 (50.8\%) & 57,226 (49.45\%)& 52,569 (46.27\%)\\
		\hline
	\end{tabular}
	\caption {Result of running BMix, WavClusteR and PARalyzer on the HUR PAR-CLIP dataset with two replicates. The number of binding sites, overlap with BMix, their occurrence in 3'UTR and introns and enrichment in RNA recognition elements (RRE) are reported, as well as the execution time of each method. The overlap with BMix equals to the percentage of binding sites reported by BMix that overlap by at least one nucleotide with the binding sites of other methods. The last line presents an analysis of reproducibility of the three different methods by assessing the percentage of RRE-enriched common sites reported by each method in both replicates.}
	\label{tableHUR}
\end{center}
\end {table}
\newpage

\begin {table}[h!]
\begin{center}
	\begin{tabular}{  l   l  c  c  c}
		~ & ~ & 1 mismatch & 2 mismatches & 3 mismatches \\ 
		\hline
		~ & No. sites & 15,317 & 21,607 & 25,289 \\
		BMix & No. miRNAs & 1024 & 1090 & 1094 \\
		~ & Intersection & 100\% & 60.13 \% & 51.12\% \\
		\hline
		~ & No. sites & 19,248 & 40,522 & 51,029 \\
		PARalyzer & No. miRNAs & 1059 & 1160 & 1160 \\
		~ & Intersection & 100\% & 45.59\% \% & 36.14\% \\
	\end{tabular}
	\caption {Results of running BMix and PARalyzer on an AGO2 PAR-CLIP dataset whose reads were aligned using Bowtie allowing for one, two or three mismatches. The number of binding sites obtained with each method in each alignment scenario is reported, as well as the numbered of covered miRNAs for each case. The intersection equals to the percentage of reported sites, obtained when two or three mismatches are used, overlapping with the binding sites obtained when only one alignment mismatch is used.}
	\label{tableBowtie}
\end{center}
\end {table}

\begin {table}[h!]
\begin{center}
	\begin{tabular}{  l   l  c  c  c}
		~ & ~ & 1 mismatch & 2 mismatches & 3 mismatches \\ 
		\hline
		~ & No. sites & 14,165 & 20,571 & 28,146 \\
		BMix & No. miRNAs & 1014 & 1050 & 858 \\
		~ & Intersection with Bowtie & 86.32\% & 80.78 \% & 58.5\% \\
	\end{tabular}
	\caption {Results of running BMix on an AGO2 PAR-CLIP dataset whose reads were aligned using TopHat allowing for one, two or three mismatches. The number of binding sites obtained with each method in each alignment scenario is reported, as well as the numbered of covered miRNAs for each case. The intersection with Bowtie is defined as the percentage of binding sites, obtained when Bowtie is used, overlapping the binding sites obtained when TopHat is used for the corresponding number of mismatches.}
	\label{tableBowtie}
\end{center}
\end {table}

\end{document}